\documentclass[prb,twocolumn,showpacs,preprintnumbers,amsmath,amssymb]{revtex4}
\usepackage{simplewick}

\usepackage{graphicx}
\usepackage{dcolumn}
\usepackage{bm}
\newcommand \be{\begin{eqnarray}}
\newcommand \ee{\end{eqnarray}}
\newcommand \ba{\begin{align}}
\newcommand \eea{\end{align}}
\newcommand {\ket}[1]{|#1\rangle}
\newcommand {\V}[1]{{\bf #1}}
\newcommand {\bra}[1]{\langle #1|}
\newcommand {\s}[1]{\sum\limits_{#1}}
\newcommand {\p}[1]{\V \partial_{\V #1}}

\begin{document}
           \csname @twocolumnfalse\endcsname
\title{Quasiparticle parameterization of meanfields, Galilei invariance and universal conserving response functions}
\author{K. Morawetz$^{1,2,3}$}
\affiliation{$^1$M\"unster University of Applied Sciences,
Stegerwaldstrasse 39, 48565 Steinfurt, Germany}
\affiliation{$^2$International Institute of Physics (IIP),
Av. Odilon Gomes de Lima 1722, 59078-400 Natal, Brazil
}
\affiliation{$^{3}$ Max-Planck-Institute for the Physics of Complex Systems, 01187 Dresden, Germany
}
\begin{abstract}
The general possible form of meanfield parameterization in a running frame in
terms of current, energy and density functionals are examined under the
restrictions of Galilean invariance. It is found that only two
density-dependent parameters remain which are usually condensed in a
position-dependent effective mass and the selfenergy formed by current and
mass. The position-dependent mass induces a position-dependent local current
which is identified for different nonlinear frames. In a second step the
response to an external perturbation and relaxation towards a local
equilibrium is investigated. The response function is found to be universal in
the sense that the actual parameterization of the local equilibrium does not
matter and is eliminated from the theory due to the conservation laws. The explicit form of the response with respect to density, momentum and energy is derived. The compressibility sum rule as well as the sum rule by first and third-order frequency moments are proved analytically to be fulfilled simultaneously. The results are presented for Bose- or Fermi systems in one- two and three dimensions.  
\end{abstract}
\maketitle

\section{Introduction}

Density functional theories have turned out to be very successful in
describing highly correlated systems. The strict proof shows that for
calculating the ground state it is sufficient to have a density-dependent
functional. This is based on the existence of a variational principle and a
functional to extremize. For excitations and in nonequilibrium it is not obvious
to find such a density functional since an extremal principle which would lead
to only one density-dependent functional are not available before the time
evolution of the density is known. Therefore
parameterizations of the meanfield in terms of observables like density,
momentum and energy are commonly met in the literature. Let us inspect first how
this is implemented in nuclear and solid state physics.

In nuclear physics there have been widely used density-dependent parameterizations of contact interactions originally introduced by
Skyrme \cite{S56,S59} to fit experimental binding properties. One can consider such density-dependent two-particle potentials as arising from three-particle interactions \cite{Ko75,M00}. As extensions the effective mean field has been described by current and energy-dependent terms \cite{EBGKV75,GNN92} with the help of which the time-dependent Hartree-Fock theories have been simplified \cite{EBGKV75}. The current contributions break explicitly the time-invariance and an effective mass appears \cite{HNPV96}. This velocity dependence of the Skyrme forces simulates finite-range effects \cite{GNN92}.

The crucial theoretical tool is the response function which provides as poles the collective excitations. Sometimes such collective excitations are described by collective variables and the response function is deduced by density variations of Skyrme-type potentials obeying frequency-weighted sum rules \cite{GBM90}.
This line of treatments starts from time-dependent Hartree-Fock equations to derive the response by a time-reversal broken Skyrme interaction \cite{BVA95,HNP96}. 

As a result the response has been used successfully to describe collective
excitations like giant resonances \cite{GBM90,BVA95,HNP96,HNPV96,MVFLS98} also
in multicomponent systems \cite{MWF97,FMW98,CLN97,Br99,MFU99,MFW99,MFW00}. In
fact the velocity dependence of the quasiparticle mean field induces the
appearance of multipole forces and when treated in random phase approximation
(RPA) produces multiple pairing forces \cite{CFL92}. An unwanted byproduct is
the violation of Galilean invariance which has to be reintroduced by symmetry
restoring forces and which shows a better agreement with the data on scissor modes in nuclei \cite{GKCR02}. A renormalized quasi-particle RPA has been developed which cures spurious states and which has been successfully applied to describe low-lying multipole vibrations \cite{KFGR04}. The removal of spurious center-of-the-mass motions is one of the most studied consequences of broken symmetries in nuclear matter \cite{BC01}.

The breaking of Galilean invariance for electrons in a solid can be restored by the construction of quasiparticles completing the Bloch theorem. The expansion of the crystal Bloch Hamiltonian around the band extrema leads to an effective quasiparticle energy which contains besides the parabolic band also velocity-dependent terms leading to the entrainment effect, where the momentum of a condensate is a linear combination of its own current and those of the other condensates \cite{G04}. The consequence is that the mass current does not agree with the mean momentum \cite{P01,P02} but two different masses appear \cite{ASVPPMTKD11,CAMD10}. Therefore neutral excitations due to correlations can carry momentum but no current \cite{NSOMFG01}.

The Galilean invariance imposes severe restrictions on the theory \cite{AGD63,CAMD10} and has been applied to plasmon frequency and Drude weights \cite{O98,O98a} and in doped graphene  \cite{ASVPPMTKD11} where a strong renormalization effect has been reported.
In the Landau Fermi liquid theory the demand on Galilean invariance leads to complicated restrictions on the energy functional \cite{CAMD10} which has been rather seldomly used explicitly, e.q. for transport phenomena in superconductivity \cite{Ju01}. It has been reported that the effect of velocity dependence described in Skyrme forces and which agrees with the experimental values cannot be reproduced when the particle-hole interaction is restricted to its Landau form \cite{GNN92}. Therefore here we will use a parameterization of the quasiparticle energy in terms of current and kinetic energy which completes the Galilean invariance and contains the momentum-dependent terms explicitly. There has been investigations of interacting Bose systems breaking the Galilean invariance due to the coupling with dispersion-less modes \cite{JR01}. However with an appropriate quasiparticle picture the Galilean invariance should be possible to restore.

All the above considerations require a density-dependent effective mass. Such
position-dependent masses have gained a great attention in semiconductor
literature. Its concept has been criticized on the basis of the Bargmann
theorem that Newton relative principle requires the mass to be a constant and
forbids a coherent superposition of states of different masses
\cite{Ro83}. Later it was refused \cite{KBr88} by showing that this theorem
does not apply to the band dependence of the effective mass. In fact it was
shown rigorously that the instantaneous Galilean invariance is in agreement
with the concept of position-dependent masses induced by the band structure or
by boundary conditions on the wave function due to abrupt heterojunctions
\cite{ZK83,Le95}. This problem has lead to a deep foundation of quantum
mechanics in terms of the Galilean group \cite{Le63,Le74}. 

The connection between Schroedinger equation based on deformed canonical commutation relations, a curved space and a position-dependent effective mass has been shown in \cite{QT04}. The resulting non-commutativity of the mass with the momentum operator can be circumvented with the concept of non-additive spatial displacements in the Hilbert space \cite{FAFA11}. The actual form of the correct effective Hamiltonian is subject to severe boundary conditions of Galilean invariance, hermiticity and probabilistic wave function \cite{So85,Ku11}. Exactly solvable effective-mass Schroedinger equations have demonstrated the usefulness of the position-dependent mass concept \cite{GGMP10,LOe10}. Even scattering of particles with position-dependent masses has been successfully described \cite{S12}.

With respect to these effective density-dependent Hamiltonians it is
desirable to have a theoretical treatment for the response function to an
external scalar perturbation. This response function determines the 
two-particle correlation function as Fourier transform of the structure
function and therefore any one-and two-particle 
observables including the optical response. It is therefore the preferred 
theoretical
object to learn about the interacting system. The position-dependent mass and
the density dependence will create complications compared to the normal
many-body treatment. As we will see this can be treated but with more
involved local currents and response function from which will be shown to complete
even higher order-sum rules.

The major line of improvement beyond the simple random phase approximation
(RPA) goes via the construction of local fields which describe the
modification of the restoring force due to the correlation of particles. Such
local fields screen the full effect of interaction at short distances
\cite{IKP84}. There exists a large literature \cite{KKM89} to construct such
local fields starting with the pioneering work of Hubbard \cite{H57}. A
deficiency of the simple local field factors showing negative pair correlation
functions has been repaired by Singwi et. al. \cite{STLS68} using
exchange-correlations  which leads to a local field factor in terms of the
structure function. Since the latter itself is expressed by the response
function a selfconsistency loop is required. This approximation has been used
and compared with molecular dynamical simulations \cite{S70,SSF96} and further
improved invoking the third-order frequency sum rule \cite{P65} by Pathak and
Vashishta \cite{PV73}. This describes the motion of particles inside the
correlation hole \cite{N74,K73}. The quantum versions have been discussed in
\cite{SchB93,HR87,DT99} and the difference between correlated and uncorrelated
occupation numbers show up in the difference of the corresponding kinetic energies leading to
further improvements \cite{VG73,IKP84} which are expressed by the virial
theorem in density derivatives of the pair correlation function
\cite{DA86,RA94}. These density variations has been used as an alternative to
construct expressions for the local fields \cite{N74,VG73,VS72}. A numerical
discussion of the Singwi, Tosi, Land and Sj\"olander local field corrections
can be found for plasma systems in \cite{BGB09} which shows appreciable
deviations from the simple RPA results. 

The third-order frequency sum rule plays
an important rule in a variety of applications. It was
used to locate the collective mode in small metal particles \cite{BE85} and to
calculate the optical dipole response in metal clusters \cite{RGB96}. Such
conserving calculations have been performed for small metal clusters
\cite{PPSH09} based on Bethe-Salpeter expansion schemes \cite{GNS85,PPSH09}. It
has been derived for electronic multilayers in \cite{LG99} and was used for
bilayer charged Bose liquids \cite{TT08}. The third-order frequency sum rule is especially important for 
low-dimensional layered structures \cite{G97,GL92,CG97} like the
 two-layered electron gas \cite{CD97,CD91,TD96}.

The hydrodynamic limit of the dynamical
structure factor has been computed in early times \cite{VH75} to access shear
and bulk viscosity of ionized matter.
Of special interest is also the compressibility sum rule. One can construct local fields directly from this sum rule \cite{DGG01}.
A bad surprise was the discovery that the compressibility 
and the third-order sum rule cannot be completed simultaneously by one static local field \cite{VG73,I84} since it violates the theorem of Ferrell, $d^2 E_0/d(e^2)^2\le0$. Therefore the focus was on the construction of dynamical local fields \cite{RA94,SchB93,RRWR99,RRRW00}. Unfortunately even the dynamic quantum
version of the
Singwi-Tosi-Land-Sj{\"o}lander local field cannot fulfill the
compressibility sum rule \cite{TB99}. The reason is that a single degree of
freedom like the selfenergy cannot provide this demand and in a former paper
it was shown that one can fulfill both sum rules by introducing two degrees of
freedom the selfenergy and the effective mass \cite{Ms01}. While this is impossible with static local
field corrections \cite{VG73,I84} the inclusion of the effective mass besides
the selfenergy allows to adapt these two quantities to complete both sum
rules. 
The prize to be paid is that the first-order frequency sum rule leads to the
effective mass instead of the bare mass as the theory demands if one starts from a basic Hamiltonian. One can form such an effective quasiparticle picture by the knowledge of the structure
factor at small distances from experiments or simulations \cite{PRBB85,BSA94,SSF96}.

Different schemes can be used  to obtain an effective Hamiltonian
characterized by density-dependent mass, current and energy. As we will show
in the appendix, the correct sum rules appear then with the effective
quasiparticle mass. There are two boundaries we demand on the theory. First to
complete the frequency as well as compressibility sum rules of the response
function. Second to find the same current from kinetic equations and from the
sum rule of the response function. That the latter demand turns out to be nontrivial is a fact mostly overlooked since one usually does not work in running frames. This reveals the
underlying conflict between Galilei invariance and sum rules met by different approximations.

Various phenomenological schemes for parameterizing 
 the response function to complete the sum rules exist  in the literature  
\cite{NI89a,NI89b,AABDTZ10}. This ranges from variational
approaches \cite{RA94}, to approximative parameterizations of the kinetic equation
\cite{UI80} up to recurrence relations \cite{H82,HL93}. Various requirements on the possible forms can be extracted from different limiting laws, for an overview see \cite{I82}

The other line of improvements includes collisional correlations in the
response \cite{MER70,D75,HPR93,KH94,RW98} mostly in relaxation time
approximation and imposes conservation laws
\cite{SM98,MF99,MF00,Ms01,BC10}. The extensions of this dielectric function 
first published by Mermin  \cite{MER70} has been applied to stopping
power problems \cite{AMDPA97,SM98}. The trick is to consider the relaxation
not towards a global equilibrium but to a local equilibrium. The latter one is
specified by the demand of conservation laws. We will show that in this way
a universal response function appears which is independent of the local equilibrium. 
We restrict here
to a one-component system though the generalization to multicomponent
systems is straight forward \cite{MWF97,MFW99} and considered in
different approaches \cite{IMTY85,KG90,Mor00} where some pitfalls have to be observed \cite{TM85}.

Other improvements rely on
numerical studies of Monte Carlo \cite{BSA94,OB94,MCS95,G97} or 
molecular dynamical simulations 
\cite{PRBB85,SSF96}. Solving the equation for the two-time Green function
\cite{KB00} provides an alternative way to sum higher order correlations
\cite{KM01}. Here we use the method to linearize the kinetic equation which
creates higher-order correlations in the response than used in the kinetic
equation itself.  Due to the variation of internal lines, already the
linearization of the meanfield kinetic equation leads to RPA (GW)
approximation. The Boltzmann equation due to the Born
diagram leads to a linear response which includes high order 
vortex corrections fulfilling sum rules consistently \cite{BMSP12,Mc02}. 
The systematic perturbative expansion of correlation functions provides in
principle the dynamical local fields \cite{BMSP12} which are of
interest for the conductivity in the long wavelength limit \cite{RRRW00} and
for the response in strongly disordered electron systems \cite{JKS03}. However the simultaneous fulfillment of frequency and compressibility sum rules remains a problem.
Alternative approaches use specific techniques useful for specific lower-dimensional systems like the response in fraction-quantum-Hall systems \cite{LF93}.
Here we formulate all expressions in $D=1,2,3$ dimensions such that any of the
above mentioned systems in nuclear, solid state and plasma physics can be treated.

\subsection{Overview about the paper}

The paper consist of five parts where first the notion of density-dependent
Hamiltonians is discussed which can be understood as created by Skyrme forces
or as parameterization of meanfields. This is performed first in the quasi-classical
picture in order to provide a feeling for the complexity of the demand of
Galilean invariance. The density-dependent parameters like effective mass,
current and selfenergy are inducing position-dependent currents. Therefore different frames are specified and discussed which serve
as benchmarks for the further treatment. The general response later is
formulated for any frame. The first chapter ends with the expected form of
compressibility. In the second chapter the kinetic equation for meanfields
with density-dependent effective Hamiltonians are discussed and the nontrivial
transformations between position-dependent frames are presented. This allows
to identify the corresponding nonlinear currents created by the
position-dependent effective masses. This quasi-classical treatment is
extended to a quantum kinetic treatment allowing the complete nonlocal
structure of the kinetic terms. A special attention is paid to the backflows
arising from the interaction of particles with the surrounding. In chapter IV
the response function is derived as linearization of the appropriate kinetic
equation around a local equilibrium. The actual form of the latter one turns
out to be irrelevant for the response function since the conservation laws
determine the form of response function. The explicit transformation rules for
the response function to change the nonlinear frames are derived. In chapter 5
the analytical form of the response function is presented within the most
convenient mixed frame. The local field  corrections are a physical way to
represent the extension of the response function from standard RPA
expressions. The static and the large frequency expansion of the latter one
provides the compressibility and frequency-weighted sum rules. With the help
of the explicit commutator relations in appendix A it is shown that the
response function derived here completes the compressibility and third-order frequency sum
rule simultaneously. Explicit expansion formulas are provided in appendix B for one, two , and three dimensions. 
In chapter 6 a summary and outlook can be found.

 \section{Quasiparticle picture and parameterization of meanfield}

\subsection{Building quantities and Galilean transformation}

We want to construct a quasiparticle picture, i.e. a meanfield, which describes Galilean-invariant excitations and which leads to a consistent response function in the sense
that the conservation laws are obeyed. We will consider a general form which one can derive from microscopic theories and will see that boundary conditions of mass current and Galilei invariance will restrict such forms considerably. 

In general we have three building
quantities, the density of particle, the momentum density or mass current and the kinetic energy density,
\be
n(\V q,t)&=&\sum\limits_p f(\V p,\V q,t)\nonumber\\
\V J(\V q,t)&=&\sum\limits_p \V{p} f(\V p,\V q,t)\nonumber\\
E_K(\V q,t)&=&\sum\limits_p {p^2\over 2 m_0} f(\V p,\V q,t)
\label{cons}
\ee 
with the bare mass $m_0$ in terms of the Wigner
function
\be
f(\V p,\V q,t)=\langle \V p+\frac 1 2 \V q|\hat \rho|\V p-\frac1 2 \V q\rangle.
\label{f}
\ee
For our purpose the difference between Wigner function and quasiparticle distribution does not play any role \cite{LSM97}.

Since we will derive all formulas for one, two and three dimensional systems (D=1,2,3) we understand 
\be
\sum\limits_p=\int {d^D p\over (2 \pi \hbar)^D}
\ee
and in equilibrium the distribution $f$ is the Fermi-or Bose function for Fermi-or Bose systems correspondingly.

Under Galiean-transform $r'=r+\V v t$ with velocity $\V v=\V u/m$ and  $\V p'=\V p+\V u$ these quantities should transform as
\be
n'&=&n \nonumber\\
\V J'&=& \V J+n \V u \nonumber\\
E'_K&=& E+{u^ 2\over 2 m_0} n+{\V u \cdot \V J\over m_0}
\label{galilei}
\ee
such that the only two possible Galilean-invariant forms read
\be
(\V p-\frac{\V J}{n})^2,\qquad I_2=2 m_0 E_K-{\V J^2\over n}.
\label{forms}
\ee
Any expression has to be build up from these two ingredients.

We search now for the quasiparticle energy excitation in terms of the
building quantities (\ref{cons})
\ba
\epsilon_p(J,E_K)=A(n) p^2+B(n)  {\V p\cdot \V J(n) \over n}+C {E_K(n)\over
  n}+\varepsilon_0(n).
\label{eps}
\end{align}
Since we consider later the linear response it is sufficient to have the linear terms
where the density-dependent coefficients $A,B,C$ have to be determined such
that the Galilean transform (\ref{galilei}) is respected. Further demands will be 
the conservation laws and that the corresponding response function should obey sum
rules. 
We want to see how much freedom remains for these general coefficients if
conservation laws and sum rules and Galilean invariance is completed
simultaneously. The density functional in the usual sense is represented by $\varepsilon_0[n]$.

\subsection{Momentum versus mass current}

The question is now how to construct a quasiparticle energy which is convenient enough to work out a consistent response function. In fact, the appropriate quasiparticle energy as argument of the distributions turns out to be nontrivial in equilibrium.

A simple guess $f_p=f(\epsilon_p)$ that the local quasiparticle distribution
is a function of the quasiparticle energy (\ref{eps}) leads immediately to a
contradiction. In fact the mean momentum from (\ref{eps}) would read
\be
\V J=\s p \V p f(\epsilon_p)=-\V J {B\over 2 A}
\label{7}
\ee
which would result into $B=-2 A$. This is in contradiction to the far reaching 
demand that the momentum density should be equal to the
current multiplied with the flux mass $m(n)$
\be
\V J=m \sum\limits_p {\partial \epsilon_p \over \partial \V p} f(\epsilon_p)=m (2 A+B) \V J
\label{i}
\ee
which would result into
\be
A\equiv {1 \over 2 m^*}=\frac 1 2 \left (\frac 1 m-B\right ).
\label{ab}
\ee

The only way out of this dilemma between (\ref{7}) and (\ref{i}) is to modify the actual quasiparticle energy (\ref{eps}) needed in defining the local equilibrium towards the local-frame quasiparticle energy 
\be
&&\tilde \epsilon_{\V p}=\epsilon_{\V p-{m^*\over n m } \V J}={ \left (\V p -{\V J\over n} \right )^2\over 2 m^*} -{J^2 B^2 m^*\over 2 n^2}+C {E_K\over n} +\varepsilon_0
\nonumber\\&&
\label{tilde}
\ee
and to chose
\be
f_{\V p}=f(\tilde \epsilon_{\V p}).
\label{motiv}
\ee

Then we have the desired agreement 
\be
\V J=\s p \V p f_p
=m \sum\limits_p {\partial \epsilon_p \over \partial \V p} f_p
\label{masscurrent}
\ee
and further $\sum\limits_p {\partial \tilde \epsilon \over \partial \V p}
f_p=0$. In this way the position-dependent coefficient $A=1/2m^*$  of the
quadratic momentum term is the effective mass. The coefficient
$B$ of the linear momentum term turns out to be the difference of the inverse
effective and flux masses.  

It is remarkable that the quasiparticle energy in the lab frame (\ref{eps})
cannot be the argument of the equilibrium distribution function. Instead we
have to have as argument the quasiparticle energy in the rest frame (\ref{tilde}).

\subsection{Galilean invariance}

The foregoing consideration is equivalent to the correct Galilean transformation provided we determine the coefficient $C$ suitably. In order to complete the Galilean invariance (\ref{galilei}) we have to have for the transformed distribution $f'_{\V p}=f_{\V p-\V u}$ since no other possibility completes all three transformations (\ref{galilei}) simultaneously. This translates with (\ref{motiv}) into the demand
\be
\tilde \epsilon_{\V p}(J',E')=\tilde \epsilon_{\V p-\V u}(J,E).
\label{13}
\ee 
With (\ref{tilde}) and (\ref{forms}) one derives from (\ref{13}) now
\be
{C\over m_0}=m^* B^2=B\left ({m^*\over m}-1\right )\equiv \left ({1\over m}-{1\over m^*}\right )^2 m^*
\ee
and the quasiparticle energy (\ref{eps}) becomes
\ba
\epsilon_p&={p^2\over 2 m}-B \left (\V p-{\V J\over n} \right )^2+{m_0 B^2 \over ( {1\over m} -B)}{E_K\over n}+ B {J^2\over n } +\varepsilon_0\nonumber\\
&={1\over 2 m^*}\left (\V p+m^* B {\V J\over n} \right )^2+\Sigma(n)
\label{eps1}
\end{align}
with
\be
\Sigma={m^* B^2\over2 n}\left ( 2 m_0 E_K \!-\!{J^2\over n}\right )+\varepsilon_0.
\label{Sigma}
\ee
The local-frame quasiparticle energy (\ref{tilde}) reads therefore
\be
\tilde \epsilon_p={1\over 2 m^*}\left (\V p-{\V J\over n} \right )^2+\Sigma.
\label{epstilde}
\ee

We can obtain different frames from the local frame by a suitable momentum shift 
\be
{\cal E}=\tilde \epsilon_{p-Q}.
\label{epstildee}
\ee
The values for the above lab frame $\epsilon_p=\tilde \epsilon_{p-Q}$ is realized by $\V Q=-m^* \V J/n m$ and for the mixed frame with quadratic dispersion $e_p=\tilde \epsilon_{p-Q}$ we need $\V Q=-\V J/n$.

In the former paper \cite{Ms01} the situation had been investigated where no currents are present and therefore $B=0$ or $m=m^*$ and $\Sigma=\epsilon_0$. Please note that the difference in the two masses has been recognized in the Fermi liquid theory \cite{O98} and is obviously reflecting the properties of the running frame.

Let us summarize that with two yet undetermined density-dependent constants,
the flux mass $m$ and $B$, or alternatively $m$ and $m^*$, or $m^*$ and
$\Sigma$, we can find a local
quasiparticle distribution (\ref{motiv}) such that the Galilei transformation
(\ref{galilei}) is completed and the mean momentum equals the mass current
(\ref{masscurrent}). One notes that the free mass $m_0$ does not appear anymore in the momentum-dependent terms. This is the reason why in the sum rules the effective mass appears and not the bare mass of the basic Hamiltonian as shown in appendix \ref{pert}.

\subsection{Local versus lab frame}
The notion of local-frame quasiparticle energy becomes justified if we calculate the mean energies. The mean local-frame quasiparticle reads
\ba
&\tilde E_{\rm qp}=\s p \tilde \epsilon_p f(\tilde \epsilon_p)
\nonumber\\&=
{m_0\over m}
\left ({m^*\over m}+2 {m\over m^*}-2\right ) 
\left (E_K-{J^2\over 2 n m_0}\right ) +n \varepsilon_0
\end{align}
and is Galilean invariant
$\tilde E_{\rm qp}'=\tilde E_{\rm qp}$
which shows that we are in the frame of moving quasiparticle. The mean lab-frame quasiparticle energy otherwise reads
\be
E_{\rm qp}&=&\s p \epsilon_p f(\tilde \epsilon_p)
\nonumber\\
&=&{m_0\over m}
\left ({m^*\over m}+2 {m\over m^*}-2\right ) E_K+B {J^2\over n} +n \varepsilon_0
\ee
which Galilei-transform as
\ba
E_{\rm qp}'-E_{\rm qp}={m_0 m^*\over m^2}\left ({u^ 2\over 2 m_0} n+{\V u \cdot \V J\over m_0}\right ).
\end{align}
We could reach that this mean quasiparticle energy Galilei-transforms as the
kinetic energy (\ref{galilei}) by fixing $m^* m_0=m^2$ but is not used here in this paper.

The local quasiparticle energies in different frames Galilei-transform itself as
\be
\epsilon_p'&=& \epsilon+B{\V u \cdot \V p}+ B^2 m^*\left ( {u^ 2\over 2 }+{\V J\over n} \cdot \V u \right )\nonumber\\
\tilde \epsilon_p'&=&\tilde \epsilon_{\V p-\V u}=\tilde \epsilon_p-{\V u \cdot \V p\over m^*}+{1\over m^*} \left ( {u^2\over 2 }+{\V J\over n} \cdot \V u \right )
\ee
which shows that the local excitations cannot be Galilean-invariant due to the position-dependent effective mass.

The difference between the two local quasiparticle energies in the lab and local frame read
\be
\epsilon_p-\tilde \epsilon_p={\V p\cdot \V J\over n m}+{J^2 \over 2 n^2 m} \left ({m^*\over m}-2 \right ).
\label{unterschied}
\ee

The mean momentum and kinetic energy in the lab-frame picture become
\be
\s p \V p f(\epsilon_p)&=&\left ( 1-{m^*\over m}\right ) \V J
\nonumber\\
\s p {p^2\over 2 m_0} f(\epsilon_p)&=&E_K-{m^*\over m}\left ( 2-{m^*\over m}\right ) {J^2\over 2 n m_0}
\label{oblab}
\ee
and one sees the difference to the expressions in the local frame (\ref{cons}).

\subsection{Mixed frame with quadratic dispersion}

Both the local and lab frame quasiparticle energies can be written into a quadratic dispersion by different momentum shifts
\be
e_p=\epsilon_{\V p - m^* {B\over n} \V J }=\tilde \epsilon_{\V p +{ \V J \over n} }={p^2\over 2 m^*}+\Sigma
\label{quadrat}
\ee
with the selfenergy (\ref{Sigma}).
It is most convenient to work in this mixed frame when it comes to linear response. Therefore we will try to formulate the kinetic equation next in this mixed frame and will provide transformation rules for the response function to reach other frames. 

Redefining the nonequilibrium distribution function
\be
{\rm f}_{\V p}(r,t)=f_{\V p+{\V J\over n}}(r,t)
\label{fg}
\ee
we can express the observables (\ref{cons}) by
\be
n(r,t)&=&\sum\limits_p {\rm f}_p\nonumber\\
\V J(r,t)&=&\sum\limits_p \V p f_p=\sum\limits_p \left (\V p+{\V J\over n}\right ) {\rm f}_p\nonumber\\
E_K(r,t)&=&\sum\limits_p {p^2\over 2 m_0} {\rm f}_p+{J^2\over 2 n m_0}
\label{obsg}
\ee
which means
\be
\sum\limits_p \V p {\rm f}_p&=&0\nonumber\\
\sum\limits_p {p^2\over 2 m_0} {\rm f}_p&=&E_K-{J^2\over 2 n m_0}={I_2\over 2 m_0}
\label{oblocal}
\ee
in difference to (\ref{oblab}) and (\ref{cons}).

\subsection{Compressibility in equilibrium}

From the explicit expression for the density (\ref{obsg}) we can see directly how the compressibility of the system should look like. The compressibility for noninteracting systems reads
\be
n^2 K_0=-\sum\limits_p\partial_{e_p} {\rm f}?p=  \beta \sum\limits_p {\rm f}_p(1 \mp {\rm f}_p)
\label{K0}
\ee
with the inverse temperature $\beta=1/k_B T$ and the upper sign for Fermions
and lower sign for Bosons. The compressibility for the interacting system we
can calculate directly
\be
n^2 K&=&\p \mu n=\beta \sum\limits_p {\rm f}_p(1\!\mp\! {\rm f}_p)\left [1\!-\!\partial_n \left ({p^2\over 2 m^*}\!+\!\Sigma\right ) \p \mu n\right]
\nonumber\\
&=&{K_0\over 1\!-\!{D\over 2} {\partial \ln m^*\over \partial \ln n}\!+\!n^2 \p n \Sigma K_0}.
\label{compress}
\ee
Here we have used a partial integration
\be
-\beta \sum\limits_p p^2 {\rm f}_p(1\mp {\rm f}_p)=m^*\sum\limits_p p \p p {\rm f}_p=-n D m^*
\ee
valid for any dimension $D=1,2,3$.

The form of compressibility (\ref{compress}) should also be the result of the compressibility sum rule for the density response function $\kappa_n$ which describes the induced density change due to an external potential
\be
\delta n =\kappa_n  \delta V^{\rm ext} =\kappa_n V_q \delta n^{\rm ext} .
\label{dela}
\ee 
The
polarization in turn describes the induced density variation in terms of the
induced potential
\be
\delta n =\Pi  \delta V^{\rm ind} 
\ee 
which itself is
the sum of
the external potential and the effective interaction
potential $(V_q+{\xi}_q) \delta n$ 
such that one gets
\be
\delta n &=&\Pi  [(V_q+\xi_q) \delta n + V_q \delta n^{\rm ext} ]
\nonumber\\
&=&{\Pi  \over 1-[V_q+\xi_q(\omega)] \Pi }V_q\delta n^{\rm ext}=\kappa_n V_q\delta n^{\rm ext}  
\label{chim}
\ee
which provides the relation between the response (\ref{dela}) and the polarization function.
The local field $\xi_q(\omega)$ describes the shielding of the interaction at short distances by particle correlations \cite{IKP84}.

Denoting the total local density by $\delta^{\rm loc}=\delta n+n^{\rm ext}$ we can write alternatively
\be
\delta n &=&\Pi  (V_q\delta n^{\rm loc} + \xi_q \delta n )
\nonumber\\
&=&{\Pi  \over 1-\xi_q(\omega) \Pi }V_q\delta n^{\rm loc}=\kappa_n^s V_q\delta n^{\rm loc}  
\label{chimloc}
\ee
which defines the screened response function $\kappa^s_n$.

The dielectric function $\varepsilon$ relates the local densities to the
external ones $\delta n^{\rm ext}=\varepsilon \delta n^{\rm loc}$ as it is customary in electrodynamics relating the displacement field to the electric field.
Therefore we can write the induced density change in terms of the external one as
\be
{\delta n\over \delta n^{\rm ext}}=\kappa_n V_q={1 \over \epsilon }-1
\ee
from which one has $\epsilon =1-V_q \kappa^s_n$.
The ratio of the induced density change to the local density change reads
\be
{\delta n\over \delta n^{\rm loc}}=\kappa^s_n V_q=1 -\epsilon.
\ee

The compressibility sum rule states now that 
\be
n^2K&=&\lim\limits_{q\to 0}{1\over V_q} {\rm Re} \left [\epsilon(q,0)-1 \right ] =-\lim\limits_{q\to 0} {\Pi(q,0)\over 1-\xi_q(0) \Pi (q,0)}
\nonumber\\&=&{K_0\over 1+n^2 K_0 \lim\limits_{q\to 0} \xi_q(0)}
\label{sumcomp}
\ee
where (\ref{K0}) has been used. Comparing with (\ref{compress}) the static local field should obey
\be
\lim\limits_{q\to 0} \xi_q(0)=\partial_n \Sigma -{D\over 2 n^2 K_0} {\partial \ln m^*\over \partial \ln n}.
\label{require}
\ee
This has to be fulfilled by the response function if the compressibility sum rule is obeyed.

Using the Kramers-Kronig relation we can write for (\ref{sumcomp}) alternatively
\ba
&n^2K=\lim\limits_{q\to 0}{1\over V_q} {\rm Re} \left [\epsilon(q,0)\!-\!1 \right ]=-\lim\limits_{q\to 0}{2\over \pi V_q} \int\limits_0^\infty{d\omega \over \omega}{\rm Im} \epsilon(q,\omega)
\nonumber\\
&=\lim\limits_{q\to 0}{2\over \pi} \int\limits_0^\infty{d\omega \over \omega}{\rm Im} {\Pi(q,\omega)\over 1 -\xi_q(\omega)\Pi(q,\omega)}
\end{align}
which illustrate the notion sum rule. We will prove that this sum rule is obeyed by the response function in chapter \ref{seccomp}.

\section{Kinetic theory}
 
\subsection{Quasi-classical kinetic equation}

\subsubsection{Local-frame kinetic equation}

The excitation of the system is described by the effective quasiparticle 
energy $\tilde \epsilon(p,q,t)$ in the local frame (\ref{tilde}). The idea is that the deviation of the distribution function from the global equilibrium one $f_0(\tilde \epsilon)$ is realized by a local equilibrium one $f^{\rm l.e.}(\tilde \epsilon)$ such that we have
\ba
\delta f=f\!-\!f_0=f\!-\!f^{\rm l.e.}\!+\!f^{\rm l.e.}\!-\!f_0=\delta f^{\rm l.e.}\!+\!{\partial f\over \partial \tilde\epsilon} \delta \tilde\epsilon
\label{le}
\end{align}
where for $f_0$ and $f^{\rm l.e.}$ the Fermi/Bose distribution serves as the equilibrium distribution.

Now we are going to construct the appropriate kinetic equation. From the foregoing consideration we have to obtain $f^{\rm l.e.}(\tilde \epsilon)$ as the local equilibrium solution of the kinetic equation which means
\be
\p {p} \tilde \epsilon \p {r} f^{\rm l.e.}-\p {r} \tilde \epsilon \p {p} f^{\rm l.e.}=0.
\ee
Therefore we can write a general local-frame linearized kinetic equation as
\be
{d \over d t}\delta f +\p {p} \tilde \epsilon \p {r} \delta f- \p {r} \delta \tilde \epsilon \p {p} f_0=\delta  \dot {\V \Phi}^{\rm Gal} \p p f_0
\label{kin1}
\ee
with a possible time-dependent backflow force $\dot {\V \Phi}^{\rm Gal}$ from
which we know at the moment only that it vanishes in equilibrium. It will be
specified later. The reason for this force is the position-dependent mass
and current which induce a backflow force and an entrainment 
which we name together Galilean force. 

The deviation of the quasiparticle energy from the equilibrium value should be understood as deviation of the local equilibrium one in the sense that it can be expressed in terms of the energy functional of the Landau theory $\delta \tilde \epsilon=\sum\limits_{p'} f_{p p'} \delta f_{p'}$.   
From now on we will understand all observables $\phi=n,\epsilon. J, E,...$ as local equilibrium ones and the deviation from it denoted as $\delta \phi$. 

Observing that $\p p \tilde \epsilon=(\V p-{\V J\over n})/m^*$ it is not difficult to see that from (\ref{kin1}) follows
\be
{d\over d t} \delta n=-\sum\limits_p \p p \delta \dot{\V \Phi}^{\rm Gal} F_0.
\ee
Assuming $\V \Phi^{\rm Gal}$ momentum-independent we see that the density excitation is a constant in time 
which is in agreement with the above notion of local-frame. In
the local frame $\tilde\epsilon$ we are local to the excitation and do not see a current.

The customary density balance reads
\be
\delta \dot {n}+\p {r} \cdot \delta {\V J^n}=0
\label{densbal}
\ee
where the particle current $\V J^n$ should have an appropriate relation to the mass current $\V J^n\sim \V J$ . In order to obtain this balance we have to go to an appropriate frame such that 
the corresponding distribution function and quasiparticle energies are changed according to
\be
f_{\V p}(\V r, t)&=&F_{\V P}\left (\V R,t\right )\nonumber\\
\tilde \epsilon_{\V p}(\V r,t)&=&{\cal E}_{\V P}(\V R,t)
\label{trafo}
\ee
with the new coordinates
$\V R=\V r +\int^t \V v_{\bar t} d \bar t$ and $\V P=\V p +\V Q$.
The relocation of the center-of-mass coordinate is given by the velocity $\V
v_t$. The accompanying momentum shift $\V Q$ has to be chosen
adequately since it describes the local excitation and formally the
Fourier transformation of the difference coordinates. 

There is an important difference whether we first transform and linearize then or the other way around. The difference is obviously a term $\sim \delta \V Q$. Transforming first and then linearizing, the current balance reads
\ba
&\sum\limits_p \V p \delta F_p=\sum\limits_p \V p \delta f_{p-Q}
\nonumber\\&=\sum\limits_p \V p\left [\left ( \delta f\right )_{p-Q}-\p {p_i}f_{p-Q}  \delta \V Q_i\right ]
\nonumber\\&
=\sum\limits_p (\V p+\V Q) \delta f_p+\sum\limits_p f_p \delta \V Q=\delta \V J+\V Q \delta n+n\delta \V Q
\label{way2}
\end{align}
which agrees with 
\ba
\sum\limits_p \V p \delta F_p&=\delta \sum\limits_p \V p F_p
=
\delta (\V J+n\V Q)=\delta \V J+\V Q \delta n+n\delta \V Q.
\label{way2a}
\end{align}
Otherwise, if one first linearizes and then transforms, one obtains
\ba
&\sum\limits_p \V p \delta F_p=\sum\limits_p \V p  \left ( \delta f\right )_{p-Q}=
\sum\limits_p (\V p+\V Q)\delta f_p
=\delta \V J+\V Q \delta n
\label{way1}
\end{align}
and we see that $\delta \V Q$ is absent compared to (\ref{way2}). This term describes just the induced backflow force when transformed to another frame.

In the following we will choose the procedure as first to linearize and then to transform. This has the advantage that all transformation obey a group property which can be handled conveniently.
With transforming (\ref{trafo}) after linearization, the observables (\ref{cons}) calculated with $\delta F_p=(\delta f)_{p-Q}$ are denoted with a tilde and we have ($J_q=\V q\cdot \V J, \V Q =\V q Q$)
\be
 \begin{pmatrix}
\delta \tilde n \cr \delta \tilde J_q\cr \delta \tilde E_K
\end{pmatrix}
={\cal D}_Q 
\begin{pmatrix}
\delta n\cr \delta J_q \cr \delta E_K
\end{pmatrix}
\label{obsgeneral}
\ee
with the matrix
\be
{\cal D}_Q=
\begin{pmatrix}
1&0&0
\cr q^2Q&1&0
\cr {q^2 Q^2\over 2 m_0}&{Q\over m_0}&1
\label{D}
\end{pmatrix}
\ee
obeying the group properties ${\cal D}_a{\cal D}_b={\cal D}_{a+b}$ and ${\cal D}^{-1}_Q={\cal D}_{-Q}$. The other way to first transform and then linearize would destroy these convenient properties.

In fact in the kinetic equation the difference in these two procedures vanishes as one can see by inspecting different choices from the 
kinetic equation (\ref{kin1}). If we first transform and then linearize we get 
\ba
&{\partial_t}\delta F+ \left (\p {P} {\cal E} +\V v\right ) \cdot \p {R} \delta F - \p {R} \delta {\cal E} \cdot \p {P} F_0
\nonumber\\&= \left ( \delta \dot \Phi _l-\dot {\delta {\V
    Q_l}}+\p {p_i} {\cal E} (\p {R_l} \delta \V Q_i- \p {R_i} \delta \V Q_l )\right ) \cdot \p {P_l} F_0.
\label{kin2}
\end{align}
One sees that besides the drift term modified by the velocity $\V v$, the momentum shift results into extra forces written on the right hand side. The latter ones can be simplified observing that 
\be
&&\p {p_i} {\cal E} \left (\p {R_l} \delta \V Q_i- \p {R_i} \delta \V Q_l \right ) \cdot \p {P_l} F_0
\nonumber\\
&&=-\p p {\cal E} \cdot \left [ \p p F_0 \times \left (\nabla \times \delta \V Q\right )\right ]
\nonumber\\&&
=-\left (\p p{\cal E} \times \p p F_0\right )\left (\nabla\times \delta \V Q\right )=0
\label{proof}
\ee
where we use the fact that the equilibrium distribution is $F_0=F({\cal E})$.
One sees that the extra terms arising if we first transform and then linearize
cancel out except the $\dot \delta \V Q$ term which we will absorb in the
backflow force since it is obviously a force established by the time
dependence of the shift current $\delta \V Q$. 

In the following we consider every time the procedure to first linearize and then transform and the final kinetic equation (\ref{kin2}) reads
\ba
&{\partial_t}\delta F+ \left (\p {P} {\cal E} +\V v\right ) \cdot \p {R} \delta F
 - \p {R} \delta {\cal E} \cdot \p {P} F_0
\nonumber\\
&=\left (\delta \dot {\V \Phi}^{\rm Gal}-\dot {\delta {\V
    Q}}\right )\cdot \p {P} F_0
\label{kin3}
\end{align}
where we understand $\p R \delta F=(\p R \delta f)_{p-Q}$ and similar for ${\cal E}$. It should be noted that the difference between these two pictures cancel out in the kinetic equation (\ref{kin3}) as we have seen in (\ref{proof}).

So far we have transformed the kinetic equation in an equivalent manner. This means we are still in the local frame as (\ref{kin1}). We can change the frame by taking into account the appropriate force on the quasiparticles. This is achieved by the transformation
\be
\p R \delta {\cal E}=(\p R \delta {\tilde \epsilon})_{p-Q}\rightarrow  \p R \delta ({\tilde \epsilon}_{p-Q}).
\label{trafo1}
\ee
The first equality expresses what we understood by the transformation so far. With the second replacement we change actually the picture to the corresponding frame.

In general frame
the density balance from (\ref{kin3}) with (\ref{trafo1}) takes the form 
\be
\delta \dot {n}+\V v\cdot \p {R} \delta n
-\left ({n\over m^*} \p R \delta \V Q\right )
=-\sum\limits_p \p p 
\delta \dot{\V \Phi}^{\rm Gal}
F_0
\label{balancen}
\ee
since $\delta \dot {\V Q}$ is independent of $p$. 
The term in the bracket on the left side appears only since we linearize first and transform then as outlined above.

In order to obtain the customary density balance (\ref{densbal}) we choose 
\be
\V v=-\V Q \p n \left ({n\over m^*}\right )
\label{vcurrent}
\ee
and obtain from (\ref{balancen}) exactly the density balance (\ref{densbal}) with the particle current
\be
\V J^n=-{n\over m^*} \V Q.
\label{pcurrent}
\ee

The backflow force on the right side of (\ref{balancen}) will lead to a
contribution if it is dependent on the momentum. We assume that the
appropriate frame is the one where the time dependence of the momentum shift
cancels this backflow force on the right side of (\ref{kin3}). Otherwise we
will get an additional frequency dependence and a renormalization of the
current response. This possibility we investigate later in a separate chapter
as unbalanced backflow.

It is instructive to rewrite (\ref{kin3}) explicitly as
\ba
&{\partial_t}\delta F+ \p {P} {\cal E} \cdot \p {R} \delta F
 - \p {R} \delta {\cal E} \cdot \p {P} F_0
\nonumber\\
&=\left \{
\partial_t \delta \left (
{\V \Phi}^{\rm Gal}
- m^*\V v-\V Q
\right )
\right .\nonumber\\&\left .
-\p R \delta 
\left (
\left [\V p-{\V J\over n}-\V Q\right ]\cdot \V v-{m^*\over 2 } v^2
\right )
\right \}
\cdot \p {P} F_0.
\label{kin4}
\end{align}
The right hand side is zero if the backflow force compensates the terms which is customary in standard derivations of kinetic equations.
Let us discuss the different frames now.

\subsubsection{Standard quasiparticle equation }

First we choose 
\be
\V Q=- {m^* \over m}{\V J \over n }
\label{choice}
\ee
such that the standard quasiparticle kinetic equation appears with the quasiparticle energy ${\cal E}=\tilde \epsilon_{p+{m^* J/n m}}=\epsilon_p$. The particle current (\ref{densbal}) and the velocity of quasiparticles become according to (\ref{pcurrent}) and (\ref{vcurrent})
\be
\V v_{\rm lab}=\p n\ln \left ({n\over m^*}\right ){\V J\over m},\qquad J^n={\V J\over m}.
\label{61}
\ee

Let us remark that (\ref{choice}) is only one of many possible choices to  
obey the necessary kinetic equation (\ref{kin1}). The only additional boundary is that the balance (\ref{densbal}) is resulting which translates into these compensation. Among these choices there is also a possible frame where the Galilean forces on the right side of (\ref{kin4}) show a form of Bernoulli potential which is appropriate when one considers superfluidity.

\subsubsection{Quasiparticle equation in mixed frame with quadratic dispersion}

Most conveniently we will work in a picture where the quasiparticle energy reads ${\cal E}_p=e_p=p^2/2m^*+\Sigma$ and shows a quadratic dispersion (\ref{quadrat}). One sees from (\ref{trafo}) that this is possible if we choose 
\be
\V Q=- {\V J \over n }.
\label{choiceq}
\ee
The corresponding particle current (\ref{densbal}) and the velocity of quasiparticles become according to (\ref{pcurrent}) and (\ref{vcurrent})
\be
\V v_{\rm mix}=\p n\ln \left ({n\over m^*}\right ){\V J\over m^*},\qquad J^n={\V J\over m^*}.
\label{vmixed}
\ee
This form of mean current for a position dependent mass will also be proven from the sum rules by quantum commutators in appendix (\ref{w2a}).

The kinetic equation reads with $\delta e_p=\delta \tilde \epsilon_{p+J/n}=\delta \epsilon_{p-m^* B J/n}$ and $F={\rm f}$
\ba
{\partial_t} \delta {\rm f} +\left (\V v+{\V p\over m^*}\right )&\p {R} \delta {\rm f}- \p {R} \delta e_p \p {p} {\rm f}_0
\nonumber\\&
=\left (\delta \dot {\V \Phi}^{\rm Gal}-\dot {\delta {\V
    Q}}\right )\cdot \p {P} {\rm f}_0.
\label{kin5a}
\end{align}
The right hand side vanishes if we chose again balanced backflows 
$\delta \Phi^{\rm Gal}=\delta \V Q
$.

Please note that in the lab frame the flux mass $m$ connects obviously the
mass current with the particle current (\ref{61}). In the mixed frame it is
the quasiparticle mass $m^*$ which connects both currents
(\ref{vmixed}). Consequently these masses will determine the corresponding 
first order frequency sum rule as it is shown in appendix \ref{pert}.

\subsection{Nonlocal and quantum calculation}

Now we extend the calculation towards the inhomogeneous case such that the
$q$-dependence has to be taken into account. We combine it with the quantum
calculation since in this way the inhomogeneous and quantum response is
described with the same formalism.

We start from the kinetic equation for the one-particle density operator in the quasiparticle picture
\be
\dot {\hat F}+i[\hat {\cal E}+\hat V^{\rm ext},\hat F] ={\cal I}
\label{1}
\ee
where $\epsilon=\langle p+\frac 1 2 q|\hat {\epsilon }|p-\frac 1 2 q\rangle$ is the quantum expectation value of (\ref{epstilde})  and the collision side ${\cal I}$ vanishes when integrated over the three moments of (\ref{cons}). The external potential $V^{\rm ext}$ creates a perturbation and excitation which we will calculate later.

The quasiparticle energy operator or meanfield can be represented in general as a Skyrme type of potential
\ba
\hat {\cal E}\!=-\!\nabla \tilde A_x\nabla\!-\!{1\over 2 i} (\tilde {\V B_x}\cdot \nabla\!+\!\nabla \cdot \tilde {\V B_x})\!+\!\tilde A' p^2\!+\!\tilde C {E_K\over n}\!+\!\varepsilon_0(n_q)
\end{align}
in analogy to the quasi-classical limit (\ref{eps}). We have 
\ba
{\cal E}&=\langle \V p+\frac 1 2 \V q|\hat {\epsilon }|\V p-\frac 1 2 \V q\rangle=p^2 (\tilde A'_q+\tilde A_q)-\V p\cdot (\tilde {\V B_q}+\V q \tilde A'_q)
\nonumber\\
&
+{q^2\over 4} (-\tilde A_q+\tilde A'_q)+\tilde C_q\star \left . {E_k\over n}\right |_q+\varepsilon_0[n_q]\nonumber\\
&=p^2 A_q+\V B_q \cdot \V p+C_q\star \left . {E_k\over n}\right |_q+\varepsilon_0[n_q]. 
\label{eps1q}
\end{align}
Here a simple renaming of $q$-dependent constants is used such that
the same form as in the homogeneous case (\ref{eps}) appears. The difference
is now that all constants are $q$-dependent which leads to convolutions
\be
J(\V q,\omega)=n_q\star \tilde {\V Q}_q=\sum\limits_k n_q \tilde {\V Q}_{q-k}
\ee
as Fourier transform of the spatial and time-dependent values
\be
\V J(\V R,t)&=&n(\V R,t) \tilde {\V Q}(\V R,t)
.
\ee
The quasiparticle energy and the effective mass are understood as spatial-dependent quantities due to the density dependence in the sense
\be
m^*(R)=\sum\limits_q {\rm e}^{i \V q \cdot \V R} m^*(n_q).
\label{mR}
\ee

The same arguments concerning the Galilean invariance as in the quasi-classical limit, eq. (\ref{ab}) and (\ref{eps1}),  apply now  resulting into 
\ba
A_q=\left . {1\over 2 m^*}\right |_q;\qquad B_q={1\over m}-{1\over m^*}
\label{eps2}
\end{align}
and analogously for the local-frame quasiparticle energy $\tilde \epsilon$ of (\ref{tilde}). The balance equation for the density follows directly from the trace of (\ref{1}) as
\ba
\dot n_q\!=\!i\sum\limits_{p\bar q}\!
\left (
\bra {p\!+\!{\bar q\over 2}}\hat {\cal E}\ket {p\!-\!q\!-\!{\bar q\over 2}}
\!-\!
\bra {p\!+\!q\!+\!{\bar q\over 2}}\hat {\cal E}\ket {p\!-\!{\bar q\over 2}}
\right )\!
F(p,\bar q).
\end{align}
The part $\hat {\cal E}$ of the Hamiltonian (\ref{Hamiltonian}) which contributes to the commutator
\ba
\bra {\V p_1}\V p A\V p\!+\!\frac 1 2 (\V p \V B\!+\!\V B \V p)\ket {\V p_2}
\!=\!\V p_1 \V p_2 A_{p_1\!-\!p_2}\!+\!{\V p_1\!+\!\V p_2\over 2} \V B_{p_1\!-\!p_2}
\end{align}
is responsible for the density balance
\be
\dot n_q+i \V q \left (2 \V J\star A_q+\V B_q\star n\right )=0
\ee
with $\V p=(\V p_1+\V p_2)/2$ and $\V q=\V p_2-\V p_1$.

Dependent on the choice of frames, see (\ref{tilde}) or (\ref{quadrat}), one obtains remembering $A=1/2 m^*$ the balance
\ba
&{\rm local-frame},\, \delta \tilde \epsilon_p,\, \V B_q=-2 A_q \star {\V J_q\over n}:\nonumber\\
&\dot n_q=0\nonumber\\
&{\rm lab-frame},\,  \delta \epsilon_p,\, \V B_q={\V J_q\over n} \star \left ({1\over m_q}-{1\over m^*_q}\right ):\nonumber\\
& \dot n_q+i \V q \left (\V J_q\star {1\over m_q}\right )=0\nonumber\\
&{\rm mixed-frame},\,  \delta e_p=\delta \tilde \epsilon_{p+J/n}=\delta \epsilon_{p-m^* B J/n},\, \V B_q=0:\nonumber\\
& \dot n_q+i \V q \left (\V J_q\star {1\over m^*_q}\right )=0
\label{74}
\end{align}
in agreement with the quasi-classical ones. Again we note that different masses connect the mass current with the particle current in different frames.

\subsection{Quasiparticle excitation}

Now we consider the excitation due to the external perturbation $V^{\rm ext}$
%
and linearize the quasiparticle energy according to
\be
{\cal E}={\cal E}_p \delta_q +\delta {\cal E}
\ee
with the equilibrium part corresponding to the chosen frame (\ref{tilde}), (\ref{oblab}) or (\ref{quadrat}) 
and the general excitation
\be
\delta{\cal E}&=& (V_0+V_4 {p^2\over 2 m_0}+V_3 \V p \cdot \V J + V_5 E_K+ V_6 J^2) \delta n \nonumber\\
&&+V_1
\V p \cdot \delta \V J+V_2 \delta E_K+V_7 \V J \cdot \delta \V J.
\nonumber\\&&
\label{de}
\ee
Some parameters are the same for all frames
\be
V_0&=&{d \varepsilon_0 \over d n};\,  V_2={C\over n}={m_0 m^*\over n} \left ({1\over m}-{1\over m^*}\right )^2;\, V_3={d V_1 \over d n} \nonumber\\
V_4&=& m_0 {d  \over d n}{1\over m^*};\, V_5={d V_2 \over d n};\, V_6=\frac 1 2 {d V_7 \over d n}
\label{para}
\ee 
and two are frame-specific 
\ba
&{\rm local},\, \delta {\cal E}=\delta \tilde \epsilon_p, \V Q=0:\nonumber\\
&V_1=-{1\over n m^*};\, V_7={1\over n^2 m^*}-{V_2\over n m_0}\nonumber\\
&{\rm lab},\,  \delta {\cal E}=\delta \epsilon_p, \V Q=-{m^*\over nm} \V J:\nonumber\\
& V_1={1\over n m}-{1\over n m^*};\, V_7=0\nonumber\\
&{\rm mixed},\,  \delta{\cal E}=\delta e_p=\delta \tilde \epsilon_{p+J/n}=\delta \epsilon_{p-m^* B J/n},\V Q=-{\V J \over n}
: \nonumber\\
& V_1=0;\, V_7=-{V_2\over n m_0}.
\label{paraspez}
\end{align}

Now we check under which restrictions the excitation (\ref{de}) itself is Galilean
invariant $\delta \epsilon=\delta \epsilon'$. 

Straightforward calculation of
(\ref{de}) with the help of (\ref{galilei}) shows that 
\ba
&\delta \epsilon'-\delta \epsilon=\biggl[ u^2 \left (V_1+{V_4\over 2 m_0}+ n V_3 + {V_2\over 2
    m_0}+ n {V_5\over 2 m_0}\right )\nonumber\\
&\!+\!{\bf p.u} \left ( V_1\!+\!{V_4\over m_0}\!+\!n V_3\right
)\!+\!{\bf u.J} \left ( V_3\!+\!{V_5\over m_0} \!+\!2 n V_6\!+\!V_7\right )\biggr ] \delta n
\nonumber\\
&+\left (
  V_1+{V_2\over m_0} +n V_7\right ) {\bf u.}\delta {\bf J}
\end{align}
and with (\ref{para})
\be
\delta \epsilon'-\delta \epsilon=
\delta \left (\frac 1 2 u^2 (d_2+n d_1)+\V p\cdot \V u d_2+ \V u \cdot \V J d_1\right )
\label{demand}
\ee
with the values of $d_{1}=V_1+{V_2\over m_0}+n V_7$ and $d_2=n V_1+{1\over m^*}$ given in table~\ref{tab1}.
\begin{table}
  \begin{tabular}{c|c|c|c}
    &$d_1$&$d_2$&$d_2+n d_1$
    \cr
    \hline
    {\rm local} &0&0&0
    \cr
    {\rm lab}&${(m^*-m)/ n m^2}$&${1/ m}$&${m^*/ n m^2}$
    \cr
    {\rm mixed}&0&${1/ m^*}$&${1/ m^*}$
  \end{tabular}
  \caption{The parameter for the Galilean-invariance breaking terms of the quasiparticle excitations in different frames.\label{tab1}}
\end{table}
We see that only in the local frame the excitation is Galilean invariant. In
the mixed or lab frame we could have Galilean-invariance of the excitations if
the masses $m^*$ and $m$ would be density-independent, i.e. position-independent.
Therefore the density-dependent mass destroys the Galilean-invariance of
excitations though the mean observables (\ref{cons}) remain of course Galilean
invariant as we have discussed in chapter II. 

We can express the necessary shift  (\ref{pcurrent}) and frame velocity (\ref{vcurrent}) for the corresponding frames in order to obtain the balance (\ref{densbal}) also in terms of the parameter (\ref{para}). Observing that 
\be
\p p \delta {\cal E}=\left ({\V p \over m_0} V_4+\V J V_3 \right ) \delta n +V_1 \delta \V J
\ee
we can repeat the integration of (\ref{kin3}) to obtain the balance (\ref{balancen}) but now obtaining
\be
\V J^n&=&-{n\over m^ *} \V Q=\left ({1\over m^*}+n V_1 \right ) \V J
\nonumber\\
\V v&=& \p n \ln {\left ({n\over m^ *}\right )} \V J^n.
\label{82}
\ee 
We see that $V_1$ determines the actual choice of the frames.

\subsection{Backflows}
The reason for the different occurring Galilei forces on the right side of (\ref{kin3}) or specifically (\ref{kin4}) or (\ref{kin5a}) and their compensations are the backflow. This backflow can be understood as dragged particles by a moving quasiparticle \cite{PN66} which means that it will be frame-dependent.  If one adds a quasiparticle to the system in general frame it carries a group velocity $\p p {\cal E}=\V v_p$. The total particle current we had from (\ref{kin3})
\be
\delta \V J^n&=&\sum\limits_p \p p {\cal E} \delta F+\sum\limits_p \p p \delta
{\cal E} F_p+\V v \delta n\nonumber\\
&=&\delta \V J_{\rm QP}^n+\delta \V J_{\rm c}^n+\delta \V J_{\rm v}^n.
\label{jn}
\ee
The last term describes the dragging of particles due to the frame velocity $\V v$ and reads explicitly (\ref{82})
\be
\V \delta J^n_v=\V J^n \delta \ln{\left( {n\over m^*}\right )}. 
\ee

The first two terms in (\ref{jn}) represent just the deviation from local equilibrium since we can write
\ba
\delta \V J_{\rm QP}^n+\delta \V J_{\rm c}^n
&=\sum\limits_p \p p {\cal E} \delta F-\sum\limits_p \delta {\cal E}\p p F
\nonumber\\
&\equiv
\sum\limits_p \V v_p \delta F-\sum\limits_p \V v^c \delta F
\nonumber\\
&=\sum\limits_p \p p {\cal E} \left (\delta F-\p {\cal E} F\delta {\cal
    E}\right )
=\sum\limits_p \p p {\cal E} \delta F^{\rm l.e.}
\label{drag}
\end{align}
where the drag velocity $\V v^c$ is given as in the Fermi-liquid theory with $\delta {\cal E}=\sum\limits_{p'} f_{pp'} \delta F_{p'}$ such that
\be
&&\sum\limits_p \delta {\cal E}\p p F=\sum\limits_p \p p {\cal E} \p {\cal E} F_p\delta {\cal E}
=\sum\limits_{pp'} f_{pp'} \p p {\cal E} \p {\cal E} F_p\delta F_{p'} 
\nonumber\\
&&=\sum\limits_{p'} \delta F_{p'} \sum\limits_p f_{pp'} \p p {\cal E} \p {\cal E} F_p
\equiv\sum\limits_{p'} \delta F_{p'} \V v^c_{p'}.
\ee
One sees from (\ref{drag}) that the group velocity $\V v_p$ is changed by the drag velocity $\V v^c$ which describes the flow of the other quasiparticles around.  We can consider this as the backflow since it arises from the interaction of moving quasiparticles with the surrounding media.

Therefore we call the first part of the particle current (\ref{jn}) the
quasiparticle current. With the quasiparticle energy in a general
frame (\ref{epstildee}) it takes the form
\be
\delta \V J^n_{\rm QP}={n \over m^*} \delta \left ({\V J\over n}\right ).
\ee
The second part we call backflow current which reads with (\ref{82})
\be
\delta \V J^n_{\rm c}={n \over m^*} \delta \left ({m^*\V J^n-\V J\over n} \right )={n \over m^*} \delta \left (m^* V_1 \V J\right )
.
\ee
We see that the parameter $V_1$ determines the backflow current and is given as the difference between the mass current of quasiparticles $m^* \V J^n$ and the momentum current $\V J$. 

The thorough treatment of backflows in metals can be found in \cite{Plas}.
When collisional correlations are considered the correct balance of backflows require the extended quasiparticle picture \cite{SL95,LSM97}. The backflow is intimately connected with the effect of collisional-drag \cite{H59} which induces a drag current from one layer to another layer \cite{R99}. The phonon-assisted drag is important for thermal transport in nanostructures \cite{SM02,YSYLM05}. In two-dimensional electron gases it was found that the backflow effect is dominant over three-body correlations for ground-state properties \cite{KCM93,HCPE03}.

\section{Response functions}

\subsection{Local equilibrium}
Next we consider the density, momentum and energy response functions due to an external perturbation $\delta V^{\rm ext}$.
Therefore the kinetic equation is linearized with respect to the quasiparticle
excitation and the density, current and energy response function are
calculated from a conserving kinetic equation with the same quasiparticle excitation.

The conserving relaxation time approximation means that we approximate the collision side of the kinetic equation (\ref{1}) by a relaxation towards a local equilibrium in the sense of (\ref{le}), 
\be
\dot {\hat F}+i[\hat {\cal E}+\hat V^{\rm ext},\hat F] ={\hat F^{\rm l.e.}-\hat F \over \tau}.
\label{1a}
\ee
The local equilibrium will be specified such that all three conservation laws
(\ref{cons}) are obeyed. We choose for the local equilibrium
distribution a (Fermi/Bose) function $F_0$ with three suitable parameters like e.g. mean current, temperature and chemical potential
\be
F^{\rm l.e.}({\bf p,r},t)=F_0\left ({\varepsilon_0({\bf p}-{\bf Q}({\bf r},t))-\mu({\bf r},t)\over T({\bf r},t)}\right ),
\label{2}
\ee
or alternatively the mass, selfenergy and current
\be
F^{\rm l.e.}({\bf p,r},t)=F_0\left ({({\bf p}-{\bf Q}({\bf r},t))^2\over m^*({\bf r},t)T}+{\Sigma({\bf r},t)-\mu\over T}\right ),
\label{2a}
\ee
or any other set of three parameters. 
The actual choice does not play a role since it vanishes from the theory as we will see now.

\subsection{Local equilibrium parameter from conservation laws}

The deviation of the local equilibrium distribution from equilibrium reads
\ba
\langle \V p\!+\!\frac{\V q}{2}|F^{\rm l.e.}\!\!-\!F_0|\V p\!-\!\frac{\V
  q}{2}\rangle \! =\!{F_0({\bf p}\!+\!{{\bf q}\over 2})\!-\!F_0({\bf p}-\!{{\bf q}\over 2}) \over {\cal E}({\bf p}\!+\!{{\bf q}\over 2})\!-\!{\cal E}({\bf p}\!-\!{{\bf q}\over 2})} \delta \epsilon^{\rm l.e.}
\label{linearization}
\end{align}
where the deviation of the quasiparticle energy from the local equilibrium is dependent on the moments  $\delta \epsilon^{\rm l.e.}=\delta \epsilon^{\rm l.e.}(1,\V p\cdot \V q,p^2/2m_0)$. If one uses the mean momentum, chemical potential and temperature as a set of observables (\ref{2}) one has e.q.
\be
\delta \epsilon^{\rm l.e.}\!=\!
\begin{pmatrix}
  1\cr \V p\cdot \V q\cr {p^2\over 2 m_0}
\end{pmatrix}^T
\!\!\begin{pmatrix}
  -{1\over T} & {q^2 Q\over m_0 T} & {\mu \over T^2}-{q^2 Q^2\over 2 m_0 T^2}
  \cr
  0&-{1\over m_0 T}& {Q\over m_0 T^2}
  \cr
  0&0&-{1\over T^2}
\end{pmatrix}
\!\!\begin{pmatrix}
    \delta \mu \cr\delta Q \cr \delta T\end{pmatrix}
\ee
or if one uses mass, current and selfenergy (\ref{2a}) one gets
\be
\delta \epsilon^{\rm l.e.}=
\begin{pmatrix}
  1\cr \V p\cdot \V q\cr {p^2\over 2 m_0}
\end{pmatrix}^T
\begin{pmatrix}
  -{q^2 Q\over m^* T}&-{Q q^2\over m^*}&1
  \cr
  {Q\over (m^*)^2}&
  -{1\over m^*} & 0
  \cr 
  {m_0 \over (m^*)^2}&0&0
\end{pmatrix}
\left (\begin{matrix}
    \delta m^* \cr\delta Q \cr \delta \Sigma\end{matrix}\right )
\ee
where we use $\V Q=Q\V q$. In general one can specify the deviation of the local quasiparticle energy by
\be
\delta \epsilon^{\rm l.e.}=
-\begin{pmatrix}
  1\cr \V p\cdot \V q\cr {p^2\over 2 m_0}
\end{pmatrix}^T
{\cal A}
\left (\begin{matrix}
    \delta^{\rm l.e.}_1  \cr\delta^{\rm l.e.}_2 \cr \delta^{\rm l.e.}_3 \end{matrix}\right )
\label{A}
\ee
where the matrix ${\cal A}$ is characteristic for the chosen local equilibrium parameter $\delta_i^{\rm l.e.}$. The actual form of ${\cal A}$ - and therefore  the form of local equilibrium specification - is not needed since it will be eliminated from the theory by conservation laws as follows.

The local equilibrium is determined by the requirement that the expectation values for density, momentum and energy are the same as the expectation values performed with the complete distribution $F$.
From the kinetic equation (\ref{1a}) we see that the conservation laws for density, momentum and energy are fulfilled if the corresponding expectation value of the collision side vanishes
\be
\sum\limits_p \phi (F-F^{\rm l.e.})&=&0.
\ee
Taking this into account we can express the deviation of the observables $\phi=1,{\bf p}, p^2/2m_0$ from equilibrium (\ref{le}),
with $\delta F =F-F_0=F-F^{\rm l.e.}+F^{\rm l.e.}-F_0$ as
\ba
\delta \phi({\bf q}, \omega)=\sum\limits_p \phi \delta F({\bf p,q},\omega)
=\sum\limits_p \phi (F^{\rm l.e.}-F_0).
\label{6}
\end{align}
Performing the momentum integrals in (\ref{6}) 
with the help of (\ref{linearization}) and (\ref{A}) we have
for the density, momentum and energy deviation 
\be
\delta {\cal X}=\left (\begin{matrix}\delta n\cr\delta J_q\cr \delta E\end{matrix}\right )=-{\cal G}(0)
{\cal A}
\left (\begin{matrix}
    \delta^{\rm l.e.}_1  \cr\delta^{\rm l.e.}_2 \cr \delta^{\rm l.e.}_3 \end{matrix}\right )
\label{g2}
\ee
where $J_q=\V q\cdot \V J$. The appearing
correlation functions are of the form
\be
g_{\phi}(\omega)&=&\sum\limits_p \phi {F_0({\bf p}+{{\bf q}\over 2})-F_0({\bf p}-{{\bf q}\over 2}) \over {\cal E}({\bf p}+{{\bf q}\over 2})-{\cal E}({\bf p}-{{\bf q}\over 2})-\omega-i0}
\label{g0}
\ee
and are condensed in matrix notation
\be
{\cal G}(\omega)&=&\left (
  \begin{matrix} 
    g_1         &g_{\bf pq}            & g_{\epsilon_0}         \cr
    g_{\bf p q} &g_{({\bf p q})^2}     & g_{{\bf p q}\epsilon_0} \cr
    g_{\epsilon_0}  &g_{\epsilon_0 {\bf pq}} & g_{\epsilon_0^2} 
  \end{matrix}
\right )
\nonumber\\
&=&\sum\limits_p 
\begin{pmatrix}
  1\cr \V p\cdot \V q\cr {p^2\over 2 m_0}
\end{pmatrix}
\otimes
{\Delta F_0\over \Delta {\cal E}-\omega-i0}
\begin{pmatrix}
  1\cr \V p\cdot \V q\cr {p^2\over 2 m_0}
\end{pmatrix}^T
\label{gg}
\ee
with $\Delta F_0=F_0({\bf p}+ {{\bf q}\over 2})-F_0({\bf p}- {{\bf q}\over
  2})$ and analogously for ${\cal E}$. The standard RPA Lindhard expression is
just $g_1(\omega)$. Here $\otimes$ stands for the dyadic product.

Frequently we will use the distribution function $F_p$ in different frames (\ref{trafo}) which translates into modified observables (\ref{obsgeneral}).
Therefore the general form of (\ref{g2}) in an arbitrary frame $F_p$  reads
\be
{\cal D}_Q \left (\begin{matrix}\delta n\cr\delta J_q\cr \delta E\end{matrix}\right )=-{\cal G}(0)
{\cal A}
\left (\begin{matrix}
    \delta^{\rm l.e.}_1  \cr\delta^{\rm l.e.}_2 \cr \delta^{\rm l.e.}_3 \end{matrix}\right )
\label{g2a}
\ee
and the correlation matrix (\ref{gg}) are calculated with $F$ and ${\cal E}$
in the general frame according to (\ref{trafo}). We will continue with this
general case and show up to what extent the final response function becomes
independent of the frame such that we can choose the most convenient mixed frame with quadratic dispersion (\ref{quadrat}) later.

By inverting (\ref{g2a}) we can eliminate the deviations from local quasiparticle energies $\delta_{1,2,3}^{\rm l.e.}$ in the balances of the kinetic equation (\ref{1a}) as we will perform now.

\subsection{Linear response from kinetic equation}

We linearize the kinetic equation (\ref{1a}) with the help of the general
form of excitations (\ref{de}) and work in the general frame which gives
\be
&&\delta F=
{\Delta F \over \Delta {\cal E}-\bar \omega-i0}
\left (\delta V^{\rm ext}+\delta {\cal E}\right )
\nonumber\\&&
+x \left ({\Delta F \over \Delta {\cal E}-\bar \omega-i0}-{\Delta F \over \Delta {\cal E}}\right ) \delta {\epsilon}^{\rm l.e.}+{\omega \delta (\V Q-\V \Phi^{\rm Gal} \p {p} F_p\over \Delta {\cal E}-\bar \omega-i0}
\nonumber\\&&
\label{l1}
\ee
with 
\be
x={1 \over i\tau \bar\omega},  \qquad \bar \omega=\omega-\V q\cdot \V v+{i\over \tau}.
\label{x}
\ee
The last term of (\ref{l1}) comes from the the $\dot \delta \V Q$ term in (\ref{kin3}) if not compensated. For the sake of completeness we keep this form though in the appropriate frame it is compensated by the backflow force $\dot {\V \Phi^{\rm Gal}}$.

By integrating (\ref{l1}) over the moments $1,\V p, p^2/2m_0$ and using  (\ref{A}) we get with the notation (\ref{g2}) 
\ba
&{\cal D}_Q\delta {\cal X}
\!=\!
\sum\limits_p
\begin{pmatrix}
  1\cr \V p\cdot \V q\cr {p^2\over 2 m_0}
\end{pmatrix}
{\Delta F \over \Delta {\cal E}\!-\!\bar \omega\!-\!i0}
\left (\delta V^{\rm ext}\!+\!\delta {\cal E}\right )
\nonumber\\&
\!-\!x\!
\sum\limits_p \!
\begin{pmatrix}
  1\cr \V p\cdot \V q\cr {p^2\over 2 m_0}
\end{pmatrix}
\!\!\otimes\!\!\left (
  {\Delta F \over \Delta {\cal E}\!-\!\bar \omega\!-\!i0}\!-\!{\Delta F \over \Delta {\cal E}}
\right)\!
\begin{pmatrix}
  1\cr \V p\cdot \V q\cr {p^2\over 2 m_0}
\end{pmatrix}^T
\!\!\!{\cal A}\!
\begin{pmatrix}
  \delta^{\rm l.e.}_1  \cr\delta^{\rm l.e.}_2 \cr \delta^{\rm l.e.}_3 \end{pmatrix}
\nonumber\\
&+\omega\sum\limits_p
\begin{pmatrix}
  1\cr \V p\cdot \V q\cr {p^2\over 2 m_0}
\end{pmatrix}
{\delta (\V Q -\V \Phi^{\rm Gal})\p {p} F_p\over \Delta {\cal E}-\bar \omega-i0}.
\label{g1}
\end{align}
Rewriting (\ref{de}) in matrix notation
\ba
\delta {\cal E}=\begin{pmatrix}
  1\cr \V p\cdot \V q\cr {p^2\over 2 m_0}
\end{pmatrix}^T
\tilde {\cal V}\delta {\cal X}
\end{align}
with the interaction matrix
\ba
\tilde {\cal V}=
\begin{pmatrix}
  V_0\!+\!V_6 {J_q^2/ q^2}\!+\!V_5 E_K&V_7{J_q/ q^2}&V_2
  \cr 
  V_3 J_q/q^2&V_1/q^2&0
  \cr 
  V_4 &0&0\end{pmatrix}
\label{v}
\end{align}
and inverting (\ref{g2a}) to eliminate ${\cal A}$ in (\ref{g1}), the 
equation for the deviations $\delta {\cal X}$ becomes 
\ba
\kappa^{-1} \delta {\cal X}\!=\!\left [
  \left ({\cal G}^{-1}(1\!+\!x)\!-\!{\cal G}_0^{-1} x\right ){\cal D}_Q\!-\!{\cal V} 
\right ]
\delta {\cal X}
\!=\!\begin{pmatrix}1\cr 0 \cr 0\end{pmatrix} \delta V^{\rm ext}.
\label{kap}
\end{align}
The inversion of the matrices in (\ref{kap}) yields the
response tensor $\kappa$.  We see that the actual form of the deviation of observables from local equilibrium has dropped out of the theory due to the demand of energy conservation (\ref{g2a}).

The occurring interaction matrix ${\cal V}$ is given by (\ref{v}) with an
additional part added in the left upper corner if one considers 
situations of unbalanced backflow forces. This comes from the time
derivative of the right hand side of (\ref{kin3}).  Let us parameterize $\partial_t \delta(\V Q-\V \Phi^{\rm Gal})\p p F \rightarrow -i\omega \V q \p p F (a {\delta J_q\over q^2}+b{J_q\over q^2} \delta n)$ with $b(n)=\partial_n a(n)$ which results into 
\be
V_0\to V_0+\omega \left ({a\over q^2} {\delta J_q\over \delta n}+b{J_q\over q^2}\right ) \delta n.
\label{fshift}
\ee 
In case that these terms occur such that $\delta \dot {\V \Phi}^{\rm Gal}$ does not cancel $\delta \dot {\V Q}$ we will discuss the consequences in the next paragraph.

From (\ref{kap}) we see how a $\V Q$-transformation is changing the response
tensor and how it looks like in different frames. If we want to express the
correlation matrix in $f(\tilde \epsilon)$ according to the local frame, we
can reabsorb this transformation into the correlation matrix ${\cal G}$ like
\ba
\tilde {\cal G}&={\cal D}_{-Q} {\cal G}(p,F_p,p) {\cal D}_{-Q}^T={\cal G}(p-Q,F_p,p-Q)\nonumber\\
&={\cal G}(p,f_p,p)
\nonumber\\
&=
\sum\limits_p 
\begin{pmatrix}
  1\cr \V p\cdot \V q\cr {p^2\over 2 m_0}
\end{pmatrix}
\otimes 
{\Delta f_p\over {{\V p \V q}\over m^*}-\V q\cdot {\V J\over n m^*}-\bar \omega-i0}
\begin{pmatrix}
  1\cr \V p\cdot \V q\cr {p^2\over 2 m_0}
\end{pmatrix}^T
\end{align}
where we used
\be
{\cal D}_{Q}\begin{pmatrix}
  1\cr \V p\cdot \V q\cr {p^2\over 2 m_0}
\end{pmatrix}
=
\begin{pmatrix}
  1\cr (\V p+\V Q)\cdot \V q\cr {(\V p+\V Q)^2\over 2 m_0}.
\end{pmatrix}.
\ee

The equation for the deviations (\ref{kap}) can be multiplied with ${\cal D}_Q^T$ from the left to yield
\be
&&\left \{
  (1+x) \tilde {\cal G}^{-1}-x\tilde {\cal G}_0^{-1}
  -{\cal D}_Q^T{\cal V}
\right \}\delta{\cal X}
\nonumber\\
&&={\cal D}_{Q}^T\begin{pmatrix}1\cr 0 \cr 0\end{pmatrix} \delta V^{\rm ext}
=\begin{pmatrix}1\cr 0 \cr 0\end{pmatrix} \delta V^{\rm ext}
\ee
and the response tensor (\ref{kap}) gets the structure
\be
\kappa^{-1}={\cal P}^{-1}-{\cal D}_{Q}^T{\cal V}
\label{113}
\ee
where the polarization tensor describes the response without meanfield ${\cal V}$
\be
{\cal P}^{-1}(\omega)=
(1+x) \tilde {\cal G}(\omega)^{-1}-x\tilde {\cal G}(0)^{-1}
\ee
and is obviously frame-independent. On contrast, the interaction matrix is multiplied by ${\cal D}^T_Q$ according to the desired frame.

Summarizing we have seen that the parameters $\delta_i^{\rm l.e.}$ of the local equilibrium distribution have been eliminated from the response function with the help of the conservation laws. This is remarkable since it shows that the response function is independent of the choice of the local equilibrium parameters and entirely determined by the conservation laws which justifies to call it universal.

\subsection{Renormalization by un-compensated backflow forces}

In case that we work in a frame where the backflow force on the right hand side of (\ref{kin3}) is not compensated we obtain an additional frequency part (\ref{fshift}) in the interaction matrix (\ref{v}). Besides the trivial shift $\omega b J_q/q^2$ in $V_0$ there appears an additional part of the current response $\delta J_q/\delta n$ with the for-factor $\tilde a=a\omega/q^2$. The latter one leads to a renormalization of the response tensor as follows. We write
\be
\left [\kappa^{-1}-
  \begin{pmatrix}\tilde a {\delta J_q \over \delta n} &0&0\cr0&0&0\cr 0&0&0
  \end{pmatrix} 
\right ]\delta {\cal X}=\begin{pmatrix}1\cr 0 \cr 0\end{pmatrix} \delta V^{\rm ext}
\ee
which means by multiplying with $\kappa$
\ba
\left [1-
  \kappa \begin{pmatrix}\tilde a{\delta J_q \over \delta n} &0&0\cr0&0&0\cr 0&0&0
  \end{pmatrix} 
  \right ]\delta {\cal X}=
\begin{pmatrix}
  1&-\tilde a \kappa_{11}&0\cr 0&1-\tilde a\kappa_{21}&0\cr0&-\tilde a \kappa_{31}&1
\end{pmatrix}
\delta {\cal X}
\end{align}
and finally
\be
\delta {\cal X}={1\over 1- a {\omega \over q^2} \kappa_{21}}
\begin{pmatrix} \kappa_{11}\cr \kappa_{21}\cr \kappa_{31}
\end{pmatrix}
\delta V^{\rm ext}.
\label{renorm}
\ee
We see that an additional renormalization of the response tensor appears
by an expression given in terms of the current response $\kappa_{21}$.

\section{Results for the response function}

\subsection{Explicit forms of correlation functions}

Let us inspect the different correlation function (\ref{g0}) as they appear in (\ref{gg}). In general frame we have 
\ba
\Delta {\cal E}_{p+{q \over 2}}-{\cal E}_{p-{q\over 2}}={\V p \V q\over m^*}+O;\,\, O =-{J_q\over n m^*}-{q^2 Q^2\over m^*}
\end{align}
with $\V Q=\V q Q$.
In fact the various correlation functions can be reduced to only three different ones given by moments of $\phi=1,p^2,p^4$ in (\ref{g0}).
One has with $\bar \omega =\omega -\V q \cdot \V v+i/\tau$
\be
g_{\V p \V q}&=&m^*\sum\limits_p \Delta F-(O-\bar \omega) m^*\sum\limits_p{\Delta F\over {\V p\V q\over m^*}+O -\bar \omega}
\nonumber\\
&=&m^*(\bar  \omega -O) g_1
\nonumber\\
g_{{p^2\V p \V q\over 2 m^*}}&=&\frac 1 2 \sum\limits_p p^2 \Delta F+{\bar \omega-O\over 2}\sum\limits_pp^2 {\Delta F\over {\V p\V q\over m^*}+O -\bar \omega}
\nonumber\\
&=&\V q \sum\limits_p \V p F+m^*(\bar \omega -O) g_{p^2\over 2 m^*}
\nonumber\\
&=&-n m^* O+m^*(\bar \omega -O) g_{p^2\over 2 m^*}
\nonumber\\
g_{(\V p \V q)^2}&=&(m^*)^2\sum\limits_p\left ( {\V p \V q\over m^*}\!-\!O\!+\!\bar \omega\right )\Delta F\!+\!(m^*)^2(O-\bar \omega)^2g_1
\nonumber\\
&=&-n q^2 m^*+(m^*)^2(O-\bar \omega)^2g_1.
\label{gexpl}
\ee
The needed moments of $(p^2/2 m_0)^n$ can be easily obtained by an appropriate
scaling of the above expressions with $(m^*/m_0)^n$. We see that the $\V Q$
transformation appearing in $O$ can be absorbed in the frequency shift except
for $g_{pq\epsilon}$ where it appears explicitly. This renders the response
formulas somewhat involved and in-transparent. We restrict ourself therefore
from now on to the mixed frame with quadratic dispersion (\ref{quadrat}) which
provides $Q=-J_q/nq^2$ due to (\ref{choiceq}) and conveniently $O=0$. Therefore the correlation functions are to be calculated with $F_p={\rm f}(e_p)$ leading to (\ref{obsg}) and (\ref{oblocal}).

\subsection{Explicit form of response function in mixed frame with quadratic dispersion}
The polarization matrix is symmetric and has the following terms $[\bar \omega=\omega-\V v\cdot \V q+i/\tau]=\Omega+i/\tau$
\ba
P_{12}&=m^* \Omega P_{11}; P_{13}=P_h P_{11};P_{23}=P_h P_{12};\nonumber\\
P_{22}&=(m^*)^2\Omega^2 P_{11}-m^* n q^2; \nonumber\\
P_{33}&=P_{11} P_h^2+(i\Omega \tau-1){g_h\over 4 m_0^2 } {g_{p^2}^2(0)-g_1(0)g_{p^4}(0) 
  \over g_1(0) g_h-i \Omega \tau g_1(\bar \omega)};\nonumber\\
P_{11}&={g_1(\bar\omega)\over 1\!-\!{1\over 1\!-\!i\Omega \tau}[1\!-\!g_s\!+\!2 m_0 P_h g_t]\!-\!{i m^* \Omega \over n q^2  \tau}g_1(\bar\omega)};\nonumber\\ 
P_h&={1\over 2 m_0} {g_{p^2}(0) g_h-i\Omega \tau g_{p^2}(\bar \omega)\over
  g_{1}(0) g_h-i\Omega\tau g_{1}(\bar \omega) }
\label{polform}
\end{align}
with the auxiliary quantities
\be
g_h&=&{g_{p^2}(\bar \omega)^2-g_{p^4}(\bar \omega)g_{1}(\bar \omega)\over g_{p^2}(0)^2-g_{p^4}(0)g_{1}(0)}  
\nonumber\\
g_s&=&{g_{p^2}(\bar \omega) g_{p^2}(0)-g_{p^4}(0)g_{1}(\bar \omega)\over g_{p^2}(0)^2-g_{p^4}(0)g_{1}(0)}  
\nonumber\\
g_t&=&{g_{p^2}(\bar \omega) g_{1}(0)-g_{p^2}(0)g_{1}(\bar \omega)\over g_{p^2}(0)^2-g_{p^4}(0)g_{1}(0)}. 
\ee

These results are the main result of the chapter and represent the universal response function in the sense that the actual form of the local equilibrium has dropped out of the theory provided the conservation laws are enforced. 

Let us compare now to
known special cases.
We have included momentum, energy and density conservation $\Pi^{\rm n,j,E}=P_{11}$. The inclusion of momentum conservation leads to the same local field correction  irrespective whether one has only density or also energy conservation considered \cite{MF00}
\be
{1\over \Pi^{\rm n,j,E}(\omega)}-{1\over \Pi^{\rm
    n,E}(\omega)}
&=&{1\over \Pi^{\rm n,j}(\omega)}-{1\over \Pi^{\rm n}(\omega)}=
- {i \omega \over \tau} {m^*\over n q^2}.
\nonumber\\&&
\label{result}
\ee
If we would have considered only density conservation the Mermin-Das polarization reads \cite{MER70,D75}
\be
\Pi^n&=&{g_1(\bar \omega)\over 1+x-x{g_1(\bar \omega)\over g_1(0)}}
={g_1(\bar \omega)\over 1+{1\over 1-i \Omega \tau}\left [
    {g_1(\bar \omega)\over g_1(0)}-1\right ]}
\ee
with (\ref{x}).

It was found that the low frequency limit of the polarization including all three conservation laws approaches the Mermin-Das (density) formula while the high
frequency limit falls with $\omega^{-5}$ compared to 
$\omega^{-3}$ for Mermin-Das polarization \cite{MF00}. The long wave length expansion of the
expression including momentum conservations shows an
excellent agreement with the complete expression for
both the high and low frequency limit. 
The corrections of order $q^2$ drop out and it is effectively of the order $q^4$.

The polarization $P_{11}$ in different notations has been discussed and compared with the Mermin-Das
dielectric function in \cite{MF99,BC10}
and has been applied to stopping power problems in plasma and storage rings \cite{SM98}. The extension to multicomponent systems has lead to a prediction of a low-lying collective mode in nuclear matter \cite{MWF97}. 

It is also instructive to write the static limit of the response functions where one has to keep care of $\lim\limits_{\omega\to 0}\bar \omega=i/\tau$. One obtains from (\ref{polform})
\be
{\cal P}(0)=  \begin{pmatrix} g_1(0) &0&{g_{p^2}(0)\over 2 m_0}\cr0&-n m^* q^2&0\cr {g_{p^2}(0)\over 2 m_0}&0&{g_{p^4}(0)\over 4 m_0^2}
  \end{pmatrix}. 
\ee
As in the case of the Mermin-Das polarization function all effects of the relaxation time vanish in the static limit.

\subsection{Density and energy response function}

With the help of the results of the foregoing chapter from (\ref{kap}) the current and energy response functions are related to the density response function in the mixed frame via
\ba
&{\delta J_q\over \delta n}= 
{J_q\over n}+m^* (\omega-\V v\cdot \V q)
\label{corr2}
\\
&{\delta E_K\over \delta n}\!=\!P_h\!+\!\tilde V_4 (P_{33}\!-\!P_{11} P_h^2)\!+\!{J_q\over n q^2} 
\!\left [\!
  {J_q\over 2  n m_0}\!+\!{m^*\over m_0} (\omega\!-\V v\cdot \V q) 
  \!\right ].
\label{corr2a}
\end{align}
Remembering the velocities (\ref{vmixed}) we see that the current response (\ref{corr2}) 
can be rewritten as
\be
-\omega \delta n+\V q \delta \left ({\V J\over m^*}\right )=0
\ee
which is exactly the density balance (\ref{densbal}) with the particle current (\ref{vmixed}). This consistency is satisfying and justifies the somewhat lengthy discussion of the introductory chapters. 

For later use we express the density derivative of the Galilean-invariant form (\ref{forms})
with the help of (\ref{corr2a}) as
\be
{\partial_n I_2\over 2 m_0}&=&\partial_n \left (E_K-{\V J^ 2\over n m_0}\right )=P_h+V_4 (P_{33}-P_{11}P_h^2)
\nonumber\\
&=& \left ( \frac 2 D+1\right ) {I_2\over 2 n m_0}+{q^2\over 8 m_0}+o\left ({1\over \omega^2}\right )
\label{I2deriv}
\ee
where the expansion formulas of (\ref{exppol}) have been used.

Finally, let us remark that if we would use the renormalization (\ref{renorm}) with the expression for the local frame $a=1/n$ we would obtain a vanishing density response $\delta n=0$ in agreement with the notion of local frame.

\subsection{Local field correction}

The density-response function results into
\ba
{\kappa_n}={\delta n\over V^{\rm ext}}
=
{P_{11}\over 1\!-\!(V_0\!+\!\xi_q) P_{11}}={g_{1}(\omega) \over 1\!-\!(V_0\!+\!\xi_q\!-\!\xi_q^*) g_{1}(\omega)}
\label{kappa}
\end{align}
with the local field correction with respect to the polarization $P_{11}$
\ba
&\xi_q=\tilde V_0-V_0\!+\tilde V_4 P_h
\!+\!\tilde V_7 {\partial J_q\over \partial n}+\tilde V_2 {\partial E_K\over \partial n}
\nonumber\\
&=\tilde V_0\!-\!V_0\!+\!(\tilde V_4+\tilde V_2) P_h
\!+\!\tilde V_2
{J_q\over n q^2}\left [{J_q\over 2 n m_0}+{m^*\over m_0}(\omega-\V v\V q) \right ]
\nonumber\\
&
+\tilde V_2\tilde V_4 (P_{33}\!-\!P_{11} P_h^2)
+\tilde V_7 \left [{J_q\over n}+m^* (\omega-\V v\V q)\right ]
\label{fq}
\end{align}
where (\ref{corr2}) and (\ref{corr2a}) have been used.

Please note that there is a local field with respect to the RPA polarization $g_1(\omega)$ itself according to (\ref{polform})
\ba
\xi_q^*=&{1\over P_{11}}-{1\over g_1(\omega)}
\nonumber\\
=&{g_s(\omega)+2 m^* P_h(\omega) g_t (\omega)-i\omega\tau\over (1-i\omega \tau) g_1(\bar \omega)}-{1\over g_1(\omega)}-{i m^* \omega\over n q^2\tau}.
\nonumber\\
=&\left \{
\begin{matrix}
0+o\left (\omega^{2} \right )
\cr
-{1\over
  1-i\omega \tau} \left ({1\over \partial_\mu n}-{2 E_K\over n^2}\right
)+o(q^2)
={1\over
  1-i\omega \tau} {8 \epsilon_f \over 15 n} +o(q^4)
\end{matrix}\right .
\label{paradox}
\end{align}
with the last equality valid for zero temperature \cite{Ms01}.
The static limit of the latter one is nonzero which is no contradiction since
 the long-wave length expansion is performed while the small-frequency
limit is written in the first line. Obviously the limits of small $\Omega$ and $q$ are not interchangeable
as it is known already from RPA Lindhard form of the dielectric function.

If we chose a frame where the backflow force leads to an additional renormalization (\ref{renorm}), it would give rise to an extra local field
\be
\xi_q\rightarrow \xi_q-a{\omega \over q^2} {\partial J_q\over \partial n}.
\ee 
This completes the form of the conserving response function obeying the three conservation laws (\ref{cons}).

\subsection{Compressibility sum rule \label{seccomp}}

For the compressibility sum rule (\ref{sumcomp}) we need to proof (\ref{require}) for the static local field factor $\lim\limits_{q\to 0}\xi_q (\Omega=0)$. 
First we note that in the static limit from (\ref{polform}) follows 
$P_{11}(0)=g_1(0)$. 
Using the values for the mixed frame (\ref{paraspez}) in which we work, it is not difficult to find 
\ba
&\xi_q(0)= \partial_n \Sigma-{m^*\over 2 n} B^2\left (\partial_n I_2-{g_2(0)\over g_1(0)}\right )
\nonumber\\&
-\frac 1 2 \partial_n \left ({1\over m^*}\right )\left [ {m^*\over 2 n}B^2\left ({g_2^2(0)\over g_1(0)}-g_4(0)\right )-{g_2(0)\over g_1(0)}\right ]
\label{fq1}
\end{align}
where we used the abbreviation for the selfenergy (\ref{Sigma}) and the Galilei-invariant form (\ref{forms}). Using the relations in appendix~\ref{exp} particularly (\ref{g124}) we see that for any dimension $D$ the long wave expansion is
\be
&&{g_2(0)\over g_1(0)}={m^* D\over nK_0}+o(q^2)
\nonumber\\
&&{g_2^2(0)\over g_1(0)}-g_4(0)=-{(m^*D)^2\over K_0}+o(q^2)
\label{h1}
\ee
and the relation from (\ref{I2n})
\ba
\partial_n I_2={m^*\over 2}\partial_n \left ({1\over m^*}\right )\left [{m^*\over K_0} D^2-(D+2) I_2\right ]+{m^* D\over n K_0}
\label{h2}
\end{align}
holds. Introducing (\ref{h1}) and (\ref{h2}) into (\ref{fq1}) we see that exactly (\ref{require}) appears as
\be
\lim\limits_{q\to 0} \xi_q(0)=\partial_n \Sigma -{D\over 2 n^2 K_0} {\partial \ln m^*\over \partial \ln n}.
\ee
This shows that the universal response function obeys the compressibility sum rule. Next we proof that the response function completes also the first and third-order frequency sum rule.

\subsection{Frequency-weighted  sum rules}
The frequency-weighted  sum rules can be easily read off from the fact that
the response function is an analytical function in the upper half
plane and falls off with large frequencies faster than 
$1/\omega^2$ such that the compact Kramers
Kronig relation reads
\be
\int d \omega ' {\kappa_n(\omega')\over \omega'-\omega -i0}=0
\label{134}
\ee
closing the contour of integration in the upper half plane. From this one has
\be
{\rm Re} \kappa_n(\omega)&=&\int {d \omega' \over \pi} {{\rm Im}
  \kappa_n(\omega')\over \omega-\omega'}
\nonumber\\
&=&{\langle\omega\rangle\over \omega^2}+{\langle\omega^3\rangle\over \omega^4}+...
\label{defm}
\ee
with the moments
\be
\langle\omega^{2 k+1}\rangle=\int {d \omega \over \pi}  \omega^{2 k+1}{\rm Im}
  \kappa_n(\omega).
\ee

The first moments are known 
\be
\langle\omega\rangle=\int {d \omega \over \pi}  \omega{\rm Im}
  \kappa_n(\omega)={n q^2\over m^*}
\label{s}
\ee
as shown in the appendix \ref{pert}, Eq. (\ref{w1}). The mass $m_0$ appears if we start with the a Hamiltonian with quadratic dispersion and the bare mass. Here we have worked in the mixed or lab frame that the mass $m^*$ and $m$ should appear respectively. Indeed, from our response function (\ref{kappa}) we obtain the large frequency limit
with the help of the appendix \ref{exp}, see also \cite{Ms01}, for all frames
\be
\langle\omega\rangle&=&n q^2 \left ({1\over m^*}+n V_1\right ).
\label{s1}
\ee
From the definition of the parameters (\ref{paraspez}) we see that for the
mixed frame $V_1=0$ such that (\ref{s}) is completed with $m^*$ as it should. If we work in the lab frame we have $\tilde V_1={1\over n m}-{1\over n m^*}$ and the sum rule (\ref{s}) is completed with $m$ as stressed already earlier after (\ref{74}).

The higher order sum rules can be obtained from the form of response function (\ref{kappa}). Using the expansions  
of the polarization
\be
{\rm Re} P_{11}(\omega)&=&{\langle\omega\rangle_P\over \omega^2}+{\langle\omega^3\rangle_P\over \omega^4}+...
\label{pe1}
\ee
and the local field 
\be
{\rm Re} \xi_q(\omega)+V_0&=&a_0+{a_2\over \omega^2}+{a_4\over \omega^4}+...
\label{pe2}
\ee
the response function (\ref{kappa}) becomes
\be
{\rm Re} \,\kappa_n(\omega)&=&P_{11}+{a_0\langle\omega\rangle_P^2\over \omega^4}+....
\label{pe3}
\ee
We see that the large-frequency expansion of the polarization function and of the response function agree up to first-order frequency sum rule, $\langle w\rangle=\langle w\rangle_P$. The first deviation arise by the third-order frequency sum rule, i.e.
\be
\langle \omega^3\rangle=\langle\omega^3\rangle_P+a_0 \langle\omega\rangle^2_P
\label{w30}
\ee 
and is given by the zeroth order expansion $a_0$ of the local field
(\ref{pe2}). 
For the polarization we obtain  with the help of (\ref{exppol}) 
\be
\langle w^3\rangle_P={3 q^2\over (m^*)^3}\left ({q^2\over D} I_2+{(\V q\cdot \V J)^2\over n}\right )+{nq^6\over 4}
\label{w3p}
\ee
and the zeroth order expansion of the local field
\be
a_0&=&\epsilon_0+{\partial_n V_2 I_2\over 2 m_0}+{V_2+V_4\over 2 m_0}\left [ \left (\frac 2 D+1\right ) {I_2\over n}+{q^2\over 4}\right ]
\nonumber\\
&=&\partial_n \Sigma+{V_4\over 2 m_0} \partial_n I_2
\label{a0}
\ee
where we have used (\ref{I2deriv}) and the fact that the form of the selfenergy (\ref{Sigma}) reads
\be
\Sigma=\epsilon_0+{V_2 I_2\over 2 m_0}.
\ee 
As derived in appendix \ref{pert}, Eq. (\ref{a03}), we have obtained with (\ref{a0}) exactly the sum rule following from the quantum commutator relations. This completes the proof of frequency sum rules and shows that the presented response function obeys simultaneously the compressibility sum rule as well as the first two energy-weighted sum rules.

\section{Summary}

An effective density-dependent Hamiltonian is considered as it appears from
Skyrme forces or meanfields. The Galilean invariance restricts the
possibilities to an effective position-dependent mass and a density-dependent
current and selfenergy. Relations between these quantities are derived which
ensure the Galilean invariance of the theory. From kinetic theory the
accompanying currents are identified which take specific forms for different
nonlinear frames and show the effect of entrainment as the influence of the
surrounding currents to the one considered. Backflow and entrainment are
interrelated and are formulated in terms of the effective mass, current and
selfenergy of the Hamiltonian. Quasi-classical and quantum expression are 
considered.

The excitation of such system shows some specific features due to the
nonlinear density dependence which are described by the density, current and
energy responses. Assuming a relaxation towards a local equilibrium the
explicit form of these response functions are calculated. It turns out that
the demand of conservation laws renders these response functions independent of
the actual form of the local equilibrium and are therefore considered as 
universal. The transformation rule is derived which translates the response functions from one nonlocal frame to another frame. 

As a satisfying feature the current response as well as frequency-weighted sum
rules up to third order are shown to be in agreement with the above identified
nonlinear and frame-dependent currents. The compressibility sum rule are
proven to be completed simultaneously with the third-order frequency sum rule
which solves a longstanding puzzle that it was considered impossible with a
static local field correction. Here the two degrees of many-body freedom,
effective mass and selfenergy are the crucial reason for this result. The
explicit quantum commutators are calculated and shown how they establish the sum
rules.

Explicit expansion formulas are given for the long wavelength and the
high-frequency limits as well as the static limit. All treatments and explicit
formulas are presented in terms of the $D=1,2,3$ dimension parameter and are
valid therefore for Bose/Fermi systems in all three dimensions.
The here derived universal and consistent response function should be possible
to use for a wide range of applications where the many-body effects are
possible to recast into an effective mass, current, self-energy and conserving
relaxation time. Especially the density functional theories belong to this
class as special case. Since the response is explicitly given for one, two and
three dimensions it should be of interest to the physics of low dimensional
materials especially their optical properties. The numerical demand does not
exceed the one calculating the RPA Lindhard finite-temperature response
function since the universal response function is expressed by
correlation functions of moments of the RPA type.  

Finally let us remark that the explicit forms of the sum rules in terms of the
density-dependent mass, current and selfenergy allows to be compared with the
ones from the standard two-body Hamiltonian with genuine two-body
interactions. Such identification allows to deduce the effective mass, current
and selfenergy which would be an alternative way to express many-body
correlations by a simpler effective density-dependent one-particle Hamiltonian 
which was treated here. These identifications are quite straight forward but
depend on the specific features and physics one wants to focus on. Therefore it has been not written down here in general. Instead the tools
to perform such construction of effective Hamiltonians are presented which is
hoped to be helpful in solid state as well as nuclear physics problems.

\acknowledgments

The encouragement by K. N. Pathak to open up this topic once more is especially
thanked for. 
This work was supported by DFG project. The financial support by the Brazilian Ministry of Science 
and Technology is acknowledged.

\appendix

\section{Perturbation theory and frequency sum rules for 1,2,3 dimensions}\label{pert}

The external potential
$\delta V^{\rm  ext}(r,t)$
induces a time-dependent change in the Hamilton operator
\be
\delta \hat H(t)=\int dr \hat n(r,t) \delta V^{\rm ext} (r,t).
\ee
The variation of the density matrix operator $\hat \rho(t)=\hat \rho+\delta
\hat \rho(t)$ can be
found from the linearized van--Neumann equation as
\be
\delta \hat \rho(t)=-i\int \limits_{-\infty}^t [\delta \hat H,\hat \rho_0]
\ee
where it has been assumed that the perturbation is conserving
symmetries of the equilibrium Hamiltonian $[\hat H_0,\delta \hat \rho]=0$.

The variation of the density expectation value $\delta n={\rm Tr} \delta
\rho\, \hat n$ is consequently 
\be
&&\delta n(r,t)=-i \int\limits_{-\infty}^t dt' \int dr' V(r',t') \langle[\hat
n(r,t),\hat n(r',t')]\rangle.
\nonumber\\&&
\label{n1}
\ee
Since in equilibrium the commutator is only dependent on the
difference of coordinates and times we can define
\ba
\langle[\hat n(r,t),\hat n(r',t')]\rangle=\!\int\! {d\omega\over \pi} {\rm e}^{-i \omega (t\!-\!t')}\sum\limits_q {\rm e}^{i q(r\!-\!r')}{\rm Im} \kappa_n(q,\omega)
\label{im}
\end{align}
from which we obtain the Fourier transform of (\ref{n1}) to
\be
\delta n(q,\omega)&=&-V^{\rm ext}(q,\omega) \int {d {\bar \omega} \over \pi} {{\rm Im } \kappa_n
  (q,{\bar \omega})\over {\bar \omega}-\omega -i 0}
\nonumber\\
&=&V^{\rm ext}(q,\omega) \kappa_n(q,\omega)
\label{ch}
\ee
identical with (\ref{dela}). To see the last identity in (\ref{ch}) we write (\ref{134}) explicitly
\ba
\int \!d \omega ' {{\rm Re} \kappa_n(\omega')\!+\!i{\rm Im}\kappa_n(\omega')\over \omega'\!-\!\omega}\!+\!i \pi {\rm Re} \kappa_n(\omega)\!-\!\pi {\rm Im} \kappa_n(\omega)=0
\end{align}
to deduce the Kramers-Kronig relations
\be
{\rm Re}\, \kappa_n(\omega)&=&-\int {d \omega '\over \pi} {{\rm Im}\,\kappa_n(\omega')\over \omega'-\omega}
\nonumber\\
{\rm Im} \, \kappa_n(\omega)&=&\frac 1 \pi \int {d \omega ' \over \pi} {{\rm Re}\,\kappa_n(\omega')\over \omega'-\omega}
\ee
which shows the second equality of (\ref{ch}).

\subsection{Sum rules\label{Chsum}}

Inverting (\ref{im}) and applying the spatial average $\int d(r_1+r_2)/2V$ one gets
\ba
{\rm Im} \kappa_n(q,\omega)&=\frac {1}{2V} 
\int d\tau {\rm e}^{i (\omega-\V q\cdot \V v) t}\langle[\hat n(q,t),\hat n(-q,0)]\rangle
\label{im2}
\end{align}
where we wrote the mean drift velocity (\ref{vcurrent}) explicitly.
From this expression it is easy to see that the frequency sum
rules read
\ba
&\int {d \omega \over \pi} \omega^n {\rm Im} \kappa_n (q,\omega)\nonumber\\&=
\frac 1 V \int dt \int {d \omega \over 2 \pi} {\rm e}^{i\omega t} (\omega+\V q\cdot \V v)^n
\langle[\hat n(q,t),\hat n(-q,0)]\rangle.
\end{align}
The first three moments read explicitly
\ba
&\int {d \omega \over \pi} \omega {\rm Im} \kappa_n (q,\omega)=\langle\omega\rangle
\nonumber\\
&\int {d \omega \over \pi} \omega^2 {\rm Im} \kappa_n (q,\omega)=2 \V q\cdot \V v \langle\omega\rangle
\nonumber\\
&\int {d \omega \over \pi} \omega^3 {\rm Im} \kappa_n (q,\omega)=\langle\omega^3\rangle+3 (\V q\cdot \V v)^2\langle\omega\rangle
\label{moments}
\end{align}
where $\langle 1\rangle=\langle\omega^2\rangle=0$. To calculate the sum rules we have by partial integration
\ba
&\langle\omega^n\rangle=\frac 1 V \int dt \int {d \omega \over 2 \pi} {\rm e}^{i\omega t} \omega^n
\langle[\hat n(q,t),\hat n(-q,0)]\rangle
\nonumber\\&=\frac 1 V
\langle[(i \partial_t)^n \hat n(q,t)|_{t\!=\!0},\hat n(-q,0)]\rangle.
\label{n2}
\end{align}
The sum rules are therefore transformed to the problem of determining the corresponding commutators.

\subsection{Effective Hamiltonian}

We consider here only the mixed frame (\ref{quadrat}). The other frames can be written down similarly. Since the response is frame-independent in the sense that we know the transformation between the different forms (\ref{113}) we choose the most convenient mixed frame with the effective quasiparticle energy $A(R)p^2+\Sigma(R)$ where $A=1/2m^*$ and $\Sigma=\epsilon_0+V_2 I_2/2 m_0$. We consider this energy as represented by the effective Hamiltonian
\be 
\hat H=\hat {\V p} \hat A \hat {\V p}+\hat {\tilde \Sigma}
\label{a12}
\ee
which has the matrix representation
\be 
H_{12}&=& \langle \V p_1| \hat H |\V p_2 \rangle
=\V p_1\cdot \V p_2 A_{\V p_1-\V p_2}+\Sigma_{\V p_1-\V p_2}
\nonumber\\
&=&\left (p^2-{q^2\over 4}\right )A_{\V q}+\tilde \Sigma_{\V q}
\label{Hamiltonian}
\ee
in terms of the difference $\V p=(\V p_1+\V p_2)/2$ and center-of-the-mass momentum  $\V q=\V p_1-\V p_2$. Here we have used the notation (\ref{mR}). We see that the term $A_q q^2/4 $ should be absorbed in 
\be
\Sigma_q=\tilde \Sigma_q-A_q {q^2\over 4}
\label{q2term}
\ee 
in order to reproduce the quasiparticle energy.

In second quantization we represent the Hamiltonian (\ref{a12}) by creation $\hat a^+$ and annihilation operators $\hat a$
\be
\hat H=\sum\limits_{12} \hat {a}_{\V p_1}^+\hat {a}_{\V p_2} \, \left (
\V p_1\cdot \V p_2 A_{\V p_1-\V p_2}+\Sigma_{\V p_1-\V p_2}\right ).
\ee
The density matrix reads
\be
\hat f_{\V p,\V q}=a_{\V p+\V q/2}^+a_{\V p-\V q/2}
\ee
such that the thermal averaging provides the Wigner distribution function
\be
f_{\V p,\V q}=\langle \hat f_{\V p,\V q} \rangle
\ee
and the density operator reads
\be
\hat n_{\V q}=\sum\limits_{\V p} \hat f_{\V p,\V q}.
\ee
With the help of the standard commutator relations it is now easy to prove the following two commutator rules
\ba
&{\rm rule}\,\,{\rm 1}:
\nonumber\\
&\left [\sum\limits_{{\V p}\bar {\V q}}\hat f_{{\V p}, \bar {\V q}} \phi_{\V p} \bar A_{{\V q}-\bar {\V q}}, \hat H\right ]=\sum\limits_{{\V p}\bar {q} \bar {\bar {\V q}}}\hat f_{{\V p},\bar {\V q}} \bar A_{{\V q}-\bar {\bar {\V q}}}
\nonumber\\
&\!\times\!
\biggl \{
\!\left [
A_{\bar {\bar {\V q}}-\bar {\V q}}\!\left (\!p^2\!-\!{\bar {q}^2\over 4}\!+\!{\bar {\V q} \cdot \bar{\bar {\V q}}\over 2}\right )\!+\!\Sigma_{\bar {\bar {\V q}}-\bar {\V q}} 
\right ]
\!\left (\!
\phi_{{\V p}-{\bar {\V q}-\bar {\bar {\V q}}\over 2}}\!-\!\phi_{{\V p}+{\bar {\V q}-\bar {\bar {\V q}}\over 2}}\!\right )
\nonumber\\
&
+A_{\bar {\bar {\V q}}-\bar {\V q}} \bar {\bar {\V q}}\cdot {\V p} \!\left (\!
\phi_{{\V p}-{\bar {\V q}-\bar {\bar {\V q}}\over 2}}\!+\!\phi_{{\V p}+{\bar {\V q}-\bar {\bar {\V q}}\over 2}}\!\right )
\!\biggr \} 
\label{rul1}
\end{align}
and
\ba
&{\rm rule}\,\,{\rm 2}:
\nonumber\\
&\left [\hat f_{{\V p}, \bar {\V q}} \phi_{{\V p},\bar {\V q}}, \hat n_{-\V q}\right ]=\hat f_{{\V p},\bar {\V q}-q} 
\left (\!
\phi_{{\V p}-{\bar {\V q}-\bar {\bar {\V q}}\over 2},\bar {\V q}}\!-\!\phi_{{\V p}+{\bar {\V q}-\bar {\bar {\V q}}\over 2},\bar {\V q}}\!\right ).
\label{rul2}
\end{align}

Applying repeatedly rule 1, (\ref{rul1}), one finds the first three time derivatives of the density operator. The first one reads
\be
i\partial_t {\hat n_{\V q}}=\left [\hat n_{\V q},\hat H\right ]=2 \sum\limits_{{\V p} \bar {\V q}} \hat f_{{\V p},\bar {\V q}} \, {\V p}\cdot {\V q} \, A_{{\V q}-\bar {\V q}}
\label{nt1}
\ee
and the thermal averaging agrees of course with the momentum-integrated
quantum kinetic equation (Vlassov), Eq. (\ref{1}), which reads in matrix representation
\ba 
i\dot f_{{\V p},{\V q}}=\sum\limits_3\left (
H_{2,3} \, f_{{\V p_1+\V p_3\over 2},\V p_3-\V p_1}-f_{{\V p_2+\V p_3\over 2},\V
  p_2-\V p_3}\, H_{3,1} \right).
\end{align}
Multiplying with $\V p$ and integrating yields the balance for the current 
\ba
&i\partial_t {\V q}\cdot {\V J}=\sum\limits_{\bar {\V q}}\sum\limits_{\V p} f_{{\V p},\bar {\V q}} 
\biggl \{\left (q^2-{\V q}\cdot \bar {\V q}\right )\tilde\Sigma_{{\V q}-\bar {\V q}}
\nonumber\\
&+
A_{{\V q}-\bar {\V q}}
\left [
 2 ({\V p}\cdot \bar {\V q})^2+(q^2-{\V q}\cdot \bar {\V q})\left (p^2-{\bar
     {\V q}^2\over 4}+{{\V q}\cdot\bar {\V q}\over 2}\right )\right ]\biggr
\}.
\label{jq}
\end{align}

The second-order time derivative of the density operator reads 
\ba
&(i\partial_t)^2 {\hat n_{\V q}}=2 \sum\limits_{{\V p} \bar {\V q}\bar{\bar {\V q}}} \hat f_{{\V p},\bar {\V q}} A_{{\V q}-\bar {\bar {\V q}}}
\nonumber\\
&\times
\biggl\{
\left [
A_{\bar {\bar {\V q}}-\bar {\V q}}\!\left (\!p^2\!-\!{\bar {q}^2\over 4}\!+\!{\bar {\V q} \! \cdot \! \bar{\bar {\V q}}\over 2}\right )+\Sigma_{\bar {\bar {\V q}}-\bar {\V q}}\right ] {\V q}\! \cdot \! (\bar {\bar {\V q}}-\bar {\V q})
\nonumber\\
&+2 {\V p}\! \cdot \! {\V q} {\V p}\! \cdot \! \bar {\bar {\V q}}A_{\bar {\bar {\V q}}-\bar {\V q}}
\biggr \}
+2 \sum\limits_{{\V p} \bar {\V q}} \hat f_{{\V p},\bar {\V q}} \, {\V p}\! \cdot \! {\V q} \, i\partial_t A_{{\V q}-\bar {\V q}}.
\label{nt2}
\end{align}

The third-order derivative takes the form
\ba
&(i\partial_t)^3 {\hat n_{\V q}}=4 \sum\limits_{{\V p} \bar {\V q}\bar{\bar {\V q}}{\V q}'} \hat f_{{\V p},\bar {\V q}} A_{{\V q}\!-\!\bar {\bar {\V q}}} \biggl \{
\!A_{{\V q}'\!-\!\bar {\V q}}A_{\bar {\bar {\V q}}-{\V q}'}
\biggl [\! 
\left (\!p^2\!-\!{\bar {q}^2\over 4}\!+\!{\bar {\V q} \! \cdot \! \bar{\bar {\V q}}\over 2}\!\right )
\nonumber\\
&\times\!
\left (
{\V p}\! \cdot \! ({\V q}'\!-\!\bar {\V q}){\V q}\! \cdot \! (\bar {\bar {\V q}}\!-\!\bar {\V q}) 
\!+\!{\V p}\! \cdot \! \bar {\bar {\V q}} {\V q}\! \cdot \! ({\V q}'\!-\!\bar {\V q})
\!+\!{\V p}\! \cdot \! {\V q} \bar {\bar {\V q}} \! \cdot \! ({\V q}'\!-\!\bar {\V q})
\right )
\nonumber\\
&
\!+\!{\V q}'\! \cdot \! {\V p} 
\biggl (
{\V q}\! \cdot \! (\bar {\bar {\V q}}\!-\!\bar {\V q}) \left (\!p^2\!+\!{(\bar {q}\!-\!{\V q}')^2\!-\!\bar {\V q}^2\over 4}\!+\!{\bar {\V q} \! \cdot \! \bar{\bar {\V q}}\over 2}\right )
\nonumber\\
&\!+\!2 {\V p}\! \cdot \! {\V q} {\V p}\! \cdot \! \bar {\bar {\V q}}\!+\!{\V q}\! \cdot \! ({\V q}'\!-\!\bar {\V q}) \bar {\bar {\V q}}\! \cdot \! ({\V q}'\!-\!\bar {\V q})\biggr )
\biggr ]\!+\!\Sigma_{{\V q}'\!-\!\bar {\V q}}A_{\bar {\bar {\V q}}\!-\!\bar {\V q}}
\biggl [ 
\nonumber\\
&
{\V p}\! \cdot \! ({\V q}'\!-\!\bar {\V q}){\V q}\! \cdot \! (\bar {\bar {\V q}}\!-\!\bar {\V q}) 
\!+\!2 {\V p}\! \cdot \! \bar {\bar {\V q}} {\V q}\! \cdot \! ({\V q}'\!-\!\bar {\V q})
\!+\!{\V p}\! \cdot \! {\V q} \bar {\bar {\V q}} \! \cdot \! ({\V q}'\!-\!\bar {\V q})
\biggr ]
\biggl\}
\nonumber\\
&+2 \sum\limits_{{\V p} \bar {\V q}\bar{\bar {\V q}}} \hat f_{{\V p},\bar {\V q}} \biggl \{
\biggl [ 
\!\left (\!p^2\!-\!{\bar {q}^2\over 4}\!+\!{\bar {\V q} \! \cdot \! \bar{\bar {\V q}}\over 2}\right ) {\V q}\! \cdot \! (\bar {\bar {\V q}}-\bar {\V q})
\nonumber\\
&
\qquad +2 {\V p}\! \cdot \! {\V q} {\V p}\! \cdot \! \bar {\bar {\V q}}
\biggr ]
\left (A_{\bar {\bar {\V q}}-\bar {\V q}}i\partial_t A_{{\V q}-\bar {\bar {\V q}}}+i\partial_t A_{\bar {\bar {\V q}}-\bar {\V q}}A_{{\V q}-\bar {\bar {\V q}}}\right )
\nonumber\\
&\qquad +{\V q}\! \cdot \! (\bar {\bar {\V q}}-\bar {\V q})\left (\Sigma_{\bar {\bar {\V q}}-\bar {\V q}}i\partial_t A_{{\V q}-\bar {\bar {\V q}}}+i\partial_t A_{\bar {\bar {\V q}}-\bar {\V q}}A_{{\V q}-\bar {\bar {\V q}}}\right )
\biggl\}
\nonumber\\
&
+2 \sum\limits_{{\V p} \bar {\V q}\bar{\bar {\V q}}} \hat f_{{\V p},\bar {\V q}} i\partial_t A_{{\V q}-\bar {\bar {\V q}}}\biggl\{2 {\V p}\! \cdot \! {\V q} {\V p}\! \cdot \! \bar {\bar {\V q}}A_{\bar {\bar {\V q}}-\bar {\V q}}
\nonumber\\
&\qquad+
\left [
A_{\bar {\bar {\V q}}-\bar {\V q}}\!\left (\!p^2\!-\!{\bar {q}^2\over 4}\!+\!{\bar {\V q} \! \cdot \! \bar{\bar {\V q}}\over 2}\right )+\Sigma_{\bar {\bar {\V q}}-\bar {\V q}}
\right ] 
{\V q}\! \cdot \! (\bar {\bar {\V q}}-\bar {\V q})\biggr \}
\nonumber\\
&
+2 \sum\limits_{{\V p} \bar {\V q}} \hat f_{{\V p},\bar {\V q}} \, {\V p}\! \cdot \! {\V q} \, (i\partial_t)^2 A_{{\V q}-\bar {\V q}}.
\label{nt3}
\end{align}

These somewhat lengthy expressions have to be subject to rule 2 (\ref{rul2}) in order to calculate the commutators (\ref{n2}). It can be tremendously simplified if we observe that the required parts for the commutator (\ref{n2}) are the ones proportional to the volume. Therefore we expand all quantities around the homogeneous equilibrium values
\be
f_{{\V p},{\V q}}=f_{\V p}\delta_{{\V q},0}+\delta f_{{\V p},{\V q}},\quad A_{\V q}=A\delta_{{\V q},0}+\delta A_{\V q}.
\ee 
Applying rule 2, (\ref{rul2}), to the derivatives (\ref{nt1}), (\ref{nt2}) and (\ref{nt3})  we obtain for the commutators (\ref{n2}) convolution structures of the form
\be
\sum\limits_{\bar {\V q}}f_{{\V p},\bar {\V q}-\V q} A_{{\V q}-\bar {\V q}} \phi_{\bar {\V q}}&=&
f_{\V p} \phi_{\V q} A\delta_{{\V q},{\V q}}+\phi_{\V q} f_{\V p} \delta A_{\V p}+\phi_{\V q} A\delta f_{{\V p},0}
\nonumber\\
&=&\left (V A f_{\V p}\!+\!f_{\V p} \delta A_{\V p}\!+\!A\delta f_{{\V p},0}\right ) \phi_{\V q}
\ee
since $\delta_{{\V q},{\V q}}=V$ representing the volume. Higher order convolutions are analogously treated. Therefore only the lowest order expansion around the homogeneous values survives in the expressions (\ref{n2}) and therefore in the terms (\ref{nt1}), (\ref{nt2}), and (\ref{nt3}).
Finally it translates into
the limits ${\V q}'=\bar {\bar {\V q}}=\bar {\V q}=\V q$ in (\ref{nt1}),
(\ref{nt2}) and (\ref{nt3}). 

We obtain from (\ref{nt1}) the first energy-weighted sum rule 
\ba
\langle \omega\rangle=\frac 1 V \langle \left [i\partial_t n_{\V q},n_{\V q}\right ]\rangle =2 q^2 A \sum\limits_{{\V p}} f_{{\V p},0}=2 n q^2 A={n q^2\over m^*}. 
\label{w1}
\end{align} 
This is the sum rule obeyed by the response function (\ref{s}). 

The second-order weighted sum rule takes the form from (\ref{nt2})
\ba
\langle \omega^2\rangle=&\frac 1 V \langle \left [(i\partial_t)^2 n_{\V q},n_{\V q}\right ]\rangle =8 q^2 A^2 \sum\limits_{{\V p}} f_{{\V p},0} {\V q}\! \cdot \! {\V p}\nonumber\\
&+2 q^2 i\partial_t A\sum\limits_{{\V p}} f_{{\V p},0}=2 q^2 A^2 {\V q}\! \cdot \! {\V J}+2 n q^2 i\dot A. 
\label{w2}
\end{align} 
All quantities are the homogeneous ones in equilibrium. 
Since we consider the linear response we have formally the time derivatives at $t=0$ to first order in the deviations of equilibrium which are the values itself. This means
\be
i\dot A|_{t=0}&=&
\partial_n A i\dot {\delta n} |_{t=0}=2 \partial_n A \left ({\V q}\cdot \delta {\V J}|_{t=0}+{\V q}\cdot {\V J} \delta A|_{t=0}\right )
\nonumber\\
&\to& 4 A \partial_n A \, {\V q}\cdot {\V J}.
\label{Adot}
\ee  
Taking into account that the mean velocity in mixed frame is just (\ref{vmixed})
which can be written
\be
{\V q}\cdot \V v=\frac 2 n {\V q}\cdot {\V J} (A+n \partial_n A)
\label{vmixeda}
\ee
we see that (\ref{w2}) is exactly
\ba
\langle \omega^2\rangle=&2\langle w\rangle {\V q}\cdot \V v
\label{w2a}
\end{align} 
which is the required form (\ref{moments}). This proves the form of mean velocity (\ref{vmixed}) also from the sum rules.

The third-order sum rule reads
\ba
&\langle \omega^3\rangle=\frac 1 V \langle \left [(i\partial_t)^3 n_{\V q},n_{\V q}\right ]\rangle \nonumber\\
&=24 A^3 q^2 \sum\limits_{{\V p}} f_{{\V p},0} ({\V q}\cdot {\V p})^2+2 q^2 A^3 \sum\limits_{{\V p}} f_{{\V p},0}\nonumber\\
&\qquad +24 q^2 A i\dot A \sum\limits_{{\V p}} f_{{\V p},0} {\V q}\cdot {\V p}+2 q^2 i^2 \ddot A\sum\limits_{{\V p}} f_{{\V p},0}\nonumber\\
&=24 A^3 q^2 \left ( {q^2\over D} I_2+{({\V q}\cdot {\V J})^2\over n}\right )+2 n q^2 A^3 
\nonumber\\
&\qquad+24 q^2 A i\dot A \,{\V q}\cdot {\V J}+2 n q^2 i\ddot A. 
\label{w3}
\end{align} 
The first two terms are just the third order expansion of the polarization
$\langle w^3\rangle_P$ according to (\ref{w3p}). Therefore we still have to prove
\be
\langle w^3\rangle=\langle w^3\rangle_P+3 ({\V q}\cdot v)^2\langle w\rangle+a_0 \langle w\rangle^2
\ee
in order to justify the form (\ref{moments}) with (\ref{w30}).

Using (\ref{vmixeda}) we see that we have to have
\ba
a_0 \langle w\rangle^2&=24 q^2 A i\dot A \, {\V q}\cdot {\V J}+2 n q^2 i\ddot A
\nonumber\\
&-24 n q^2 A \,({\V q}\cdot {\V J})^2\left [ 2 A \partial_n A + n(\partial_n A)^2\right ].
\label{a01}
\end{align}
Since we multiply in the first term $i\dot A$ already with ${\V q}\cdot {\V
  J}$ we have to use from (\ref{Adot}) only the second term, otherwise we
would get a quadratic response at $t=0$.

The second derivatives of $A$ requires some more care. We have from (\ref{Adot})
\be
i^2\ddot A=2 i\partial_t \left [\partial_n A \left ({\V q}\cdot \delta {\V J}+{\V q}\cdot {\V J} \delta A\right )\right ].
\label{Adot2}
\ee
From all the appearing 6 explicit time-derivative terms only one remains as first order response since (\ref{Adot2}) is already  multiplied with ${\V q}\cdot {\V J}$ in (\ref{a01}),
\ba
&i^2 \ddot A=2 A \partial_n A i\partial_t {\V q}\cdot \delta {\V J}
\nonumber\\
&= 4 A \partial_n A \left \{
\sum\limits_p f_{{\V p},0} [({\V p}\cdot {\V q})^2+q^2 p^2]+{q^2\over 2} n \, \delta \tilde \Sigma
\right \}
\label{Adot2a}
\end{align}
where we used (\ref{Adot}) and linearized (\ref{jq}). Observing with
(\ref{q2term}) that $\partial_n A\, \delta \tilde \Sigma=\partial_n A \partial_n\tilde  \Sigma \delta n=\partial_n \tilde
\Sigma\delta A\to \partial_n \tilde \Sigma A=\partial_n \Sigma A+A q^2/4$ we see that we obtain after cancellation of terms
\ba
a_0=\partial_n\Sigma\!+\!\partial_n A\!\left [ \!\left (\!{2\over
      D}\!+\!1\!\right )\! q^2 I_2\!+\!{({\V q}\cdot {\V J})^2\over n}\!\right
]\!=\partial_n \Sigma\!+\!\partial_n \!A \partial_n I_2
\label{a03}
\end{align}
where we have used the identity (\ref{I2deriv}). The expression (\ref{a03}) is just the one
we have obtained from the sum rule of the response function (\ref{a0}) which completes the proof.

\section{Expansion formulas\label{exp}}
\subsection{General relations}

We provide here the expansion formulas for any dimension $D=1,2,3$ and work in the mixed frame where the distribution is $g(e_p)$ and the quasiparticle energy is $e_p=p^2/2 m^*+\Sigma$. First we observe that for any dimension we have with angular integration $d \alpha$ by partial integration
\be
\sum\limits_p p^n\partial_\epsilon {\rm f}&=&m^*\int d\alpha \int\limits_0^\infty dp p^{D+n-2}\partial_p {\rm f}
\nonumber\\
&=&-m^*(D+n-2)\sum\limits_p p^{n-2} {\rm f}.
\label{bez}
\ee
Using the definition of the compressibility, $\partial_\mu=n^2 K\partial_n$, and with the help of (\ref{bez}) one has
\be
\partial_n I_2={m^* D\over n K}\!-\!(D\!+\!2) {m^*}\partial_n \left ({1\over m^*}\right ) I_2\!-\!n m^* D \partial_n \Sigma
\label{I2n}
\ee
which we use after introducing $K_0$ from (\ref{compress}) to derive (\ref{h2}).

Next we rewrite the correlation functions $g_n$ of (\ref{g0})
\be
g_n(\omega)&=&\sum\limits_p p^n {{\rm f}({\bf p}+{{\bf q}\over 2})-{\rm f}({\bf p}-{{\bf q}\over 2}) \over {\V {\V p}\cdot \V q\over m^*}-\omega-i0}.
\label{g0b}
\ee
It is convenient to introduce $x=\V {\V p}\cdot \V q/q$ and $k=m^*\omega/q$. Then in the integral of $g_2$ we write
\be
{p^2\over x-k}=(p^2-x^2)+x+k+{k^2\over x-k}
\ee
to obtain
\be
g_2=\tilde \Pi_2+\left ({m^*\omega\over q}\right )^2 g_0-n m^*
\label{g2w}
\ee
 with the convenient form
\be
\tilde \Pi_2&=&\sum\limits_p \left (p^2-{\V {\V p}\cdot \V q \over q}\right ) {{\rm f}({\bf p}+{{\bf q}\over 2})-{\rm f}({\bf p}-{{\bf q}\over 2}) \over {\V {\V p}\cdot \V q\over m^*}-\omega-i0}
\ee
which vanishes e.q. in 1D.

Similarly we can write for $g_4$
\be
{p^4\over x\!-\!k}&=&{(p\!-\!x)^4\over x\!-\!k}\!-\!x^3\!-\!k x^2\!+\!(2 p^2\!-\!k^2) x\!+\!k (2 p^2\!-\!k^2)
\nonumber\\&&
+{k^2 (2 p^2-k^2)\over x-k}.
\ee
The different occurring integrals over the angle $x$ can be performed in any dimension $D=1,2,3$ and we find
\be
&&\sum\limits_p x^2 {\rm f}={1\over D}\sum\limits_pp^2 {\rm f},\quad
\sum\limits_p x^4 {\rm f}={3\over D(D+2)}\sum\limits_pp^4 {\rm f},
\nonumber\\
&&\sum\limits_p x^6 {\rm f}={5\over 4 D^2-D+2}\sum\limits_pp^6 {\rm f}.
\label{angular}
\ee
With the help of (\ref{angular}) one has
\ba
g_4&=\tilde \Pi_4+{2 (m^*)^2 \omega^2 \over q^2}g_2-{(m^*)^4 \omega^4 \over q^4} g_0
\nonumber\\
&-{n m^* q^2\over 4}\left (1-{4 (m^*)^2 \omega^2\over q^4}\right )-\left (\frac 1 D+2\right ) m^* I_2
\nonumber\\
&=\tilde \Pi_4+{2 (m^*)^2 \omega^2 \over q^2}\tilde\Pi_2+{(m^*)^4 \omega^4 \over q^4} g_0
\nonumber\\
&-{n m^* q^2\over 4}\left (1+{4 (m^*)^2 \omega^2\over q^4}\right )-\left (\frac 1 D+2\right ) m^* I_2.
\label{g4w}
\end{align}

\subsection{Static long wave length expansion}

In the static limit we have for (\ref{g2w}) and (\ref{g4w})
\be
g_2(0)&=&\tilde \Pi_2(0)-n m^*\nonumber\\
g_4(0)&=&\tilde \Pi_4(0) -n m^* {q^2\over 4}-\left ( 2+\frac 1 D\right ) m^* I_2.
\label{g24}
\ee
For the long-wavelengths expansion we use again $x=\V {\V p}\cdot \V q/q$ and find for the static argument of $\tilde \Pi$
\ba
{{\rm f}({\bf p}\!+\!{{\bf q}\over 2})\!-\!{\rm f}({\bf p}\!-\!{{\bf q}\over 2}) \over {\V {\V p}\cdot \V q\over m^*}}=\partial_e {\rm f}\!+\!{q^2\over 8 m^*} \partial_e^2 {\rm f}\!+\!{q^2x^2 \over 24 (m^*)^2} \partial_e^3 {\rm f}\!+\!o(q^4).
\label{gexp}
\end{align}
Using (\ref{angular}) and repeatedly (\ref{bez}) one gets
\be
\tilde \Pi_2(0)&=&-n m^* (D-1)+{n^2 q^2\over 12 } (D-1) K_0+o(q^4)
\nonumber\\
\tilde \Pi_2(0)&=&-m^* \left (D-\frac 1 D \right )I_2+o(q^2)
\ee
and for (\ref{g24}) finally
\be
g_1(0)&=&-n^2 K_0+o(q^2)
\nonumber\\
g_2(0)&=&-n m^* D+o(q^2)
\nonumber\\
g_4(0)&=&-m^2(2+D) I_2+o(q^2).
\label{g124}
\ee

\subsection{Dynamic long wave length expansion}

Expanding the denominator in $g_n$ of (\ref{g0b}) or (\ref{g0}) and using (\ref{gexp}) as well as the fact that only even exponents of $x$ count, one gets
\ba
g_n(\omega)=&-{q\over m^*\omega}\sum\limits_p p^n\left (
{q x^2\over m^*\omega}\partial_e {\rm f}+{q^3x^2\over 8 (m^*)^2\omega} \partial_e^2 {\rm f}\right .
\nonumber\\
&\left .+{q^3x^4 \over 24 (m^*)^3\omega} \partial_e^3 {\rm f}\!+\!{q^3x^4 \over (m^*)^3\omega^3} \partial_e {\rm f}\!+\!o(q^5)
\right )
\end{align}
and after using again (\ref{angular}) and repeatedly (\ref{bez}) for $n\ge 2$
\be
g_n(\omega)&=&{q^2\over m^*\omega^2} {D+n\over D} I_n+
{3q^4\over (m^*)^3\omega^4} {D+n+2\over D(D+2)} I_{n+2}
\nonumber\\
&+&
{q^4\over 8 m^*\omega^2} {(D\!+\!n\!-\!2)(D\!+\!n) n\over D(D\!+\!2)} I_{n\!-\!2}\!+\!o(q^6)
\ee
and
\be
g_0(\omega)&=&{n q^2\over m^*\omega^2}+
{3q^4\over (m^*)^3\omega^4} {1\over D} I_{2}.
\ee

\subsection{Dynamic large frequency expansion}

The expansion with respect to large frequencies works similar as the expansion
with respect to small wavelength with the difference that higher-order
wavelength enters the corresponding terms. First we observe that the form $\V
p-x \V q$ with $x={\V p}\cdot \V q/q$ is invariant under transformation $\V p\to \V p\pm \V q/2$ and therefore $p^2-x^2$ as well. This shows that we can expand in a geometric sum understood as difference of upper sign expressions minus lower ones
\ba
\tilde \Pi_n&=-\frac 1 \omega \sum\limits_p (p^2-x^2)^{n/2} {\rm f}_p
\sum\limits_-\left [
1+{q \left (x\mp\frac q 2\right )\over m\omega}  +...
\right ] 
\nonumber\\
&={q^2\over m \omega^2} \sum\limits_p (p^2\!-\!x^2)^{n/2} {\rm f}_p
\left [
1\!+\!{q^4\over 4 m^2 \omega^2} \left (1\!+\!{q^6\over 4 m^4 \omega^4}\right ) 
\right .
\nonumber\\
&
\left . +x^2 {q^2\over m^2\omega^2}\left (3+{5 q^4\over 2 m^2 \omega^2}\right )-5 x^4 {q^4\over m^4\omega^4}
\right ]+o\left (\omega^{-8}\right ).
\end{align}
This expansion is different from the long-wavelength expansion of the foregoing section.

Abbreviating $y=1/k=q/m\omega$ one obtains for the needed expansion order in $\omega$
\ba
g_0(\omega)&={q^2\over m \omega^2}\left [n\left (1+{q^2 y^2\over 4}+{q^4 y^4\over 16}\right )+{y^2 (6+5 q^2 y^2) \over 2 D } I_2
\right .
\nonumber\\
&\left .+{15 y^4\over D(2+D)} I_4\right ] +o\left (\omega^{-8}\right )
\nonumber\\
g_2(\omega)&={q^2\over 4 m \omega^2}\left [{n q^2}\left (1+{q^2 y^2\over 4}\right )+{12 (4+D) y^2\over D(2+D)} I_4
\right .
\nonumber\\
&\left .
+(4 D+8+(D+9)q^2 y^2){I_2\over D}
\right ] +o\left (\omega^{-6}\right )
\nonumber\\
g_4(\omega)&={q^2\over m \omega^2}\left [n {q^2\over 16}\!+\!{4\!+\!D\over 2 D } q^2 I_2\!+\!\left (1\!+\!{4\over D}\right )I_4\right ] \!+\!o(\omega^{-4}).
\end{align}

With the help of this expansion the polarization functions (\ref{corr2a}) expand as
\ba
&P_h=\left ( \frac 2 D+1\right ) {I_2\over 2 n m_0}+{q^2\over 8 m_0}+o\left (\omega^{-2}\right )
\nonumber\\
&P_{33}-P_{11}P_h^2
=0
\nonumber\\
&
+{q^2 \left [n q^2 I_2 D+n (4+D)D I_4
-(2+D)^2 I_2^2\right ]
\over 4 n m \omega^2 m_0^2 D^2} +o\left (\omega^{-4}\right ).
\label{exppol}
\end{align}

\bibliography{bose,kmsr,kmsr1,kmsr2,kmsr3,kmsr4,kmsr5,kmsr6,kmsr7,delay2,delay3,spin,spin1,refer,iso,gdr,chaos,sem3,sem1,sem2,short,cauchy,genn,paradox,deform,blase}

\end{document}